\documentclass[a4paper,11pt]{article}
\pdfoutput=1

\usepackage{amssymb,amsmath,bm}
\usepackage{a4wide}
\usepackage{color}
\usepackage{slashed}
\usepackage{graphicx}
\usepackage[latin1]{inputenc}
\usepackage{amsfonts}
\usepackage{lscape}
\usepackage{amsthm}
\usepackage{booktabs}
\usepackage{array}
\usepackage{rotating}
\usepackage[numbers, sort]{natbib}
\usepackage{comment}
\usepackage{float}

\makeatletter
\@addtoreset{equation}{section}
\makeatother

\usepackage[small,bf]{caption}
\usepackage{subcaption}

\newcommand{\bea}{\begin{eqnarray}}
\newcommand{\eea}{\end{eqnarray}}
\newcommand{\nn}{\nonumber}

\newcommand{\TeV}{{\rm TeV}}
\newcommand{\eps}{\epsilon}
\newcommand{\ord}[1]{\mathcal{O}\left( #1 \right)}

\newcommand{\be}{\begin{equation}}
\newcommand{\ee}{\end{equation}}
\newcommand{\vev}[1] {\langle #1 \rangle}

\newcommand{\fb}{\overline{5}}
\newcommand{\Xfb}{X_{\overline{5}}}
\newcommand{\epschi}{\epsilon_{\chi}}
\newcommand{\epsphi}{\epsilon_{\phi}}
\newcommand{\epsu}{\epsilon_u}
\newcommand{\epsd}{\epsilon_d}

\newcommand{\epst}{\epsilon_3}
\newcommand{\epsup}{\epsilon_u^\prime}

\newcommand{\mF}{\tilde{m}_F}
\newcommand{\mD}{\tilde{m}_D}

\newcommand{\sqrtd}{\sqrt{m_d/m_s}}

\begin{document}

\begin{flushright} 
CPHT-RR074.0713 \\
FLAVOUR(267104)-ERC-48\\
ICTP-SAIFR/2013-010
\end{flushright}

\thispagestyle{empty}

\vspace*{1cm}

\begin{center}

{\Large\bf Linking natural supersymmetry to flavour physics }

\vspace*{5mm}

\vspace*{5mm} 
\vskip 0.5cm

\centerline{\bf
Emilian Dudas$^a$, Gero von Gersdorff$^b$, Stefan Pokorski$^c$, Robert Ziegler$^{d}$}
\vskip 5mm
\small{$^a$ Centre de Physique Th\'eorique, \'Ecole Polytechnique, CNRS, Palaiseau, France.}\\
\vspace{0.1cm}
\small{$^b$ ICTP South American Institute for Fundamental Research, Instituto de Fisica Teorica,\\
Sao Paulo State University, Brazil} \\
\vspace{0.1cm}
\small{$^c$ Institute of Theoretical Physics, Faculty of Physics, University of Warsaw, Warsaw, Poland}\\
\vspace{0.1cm}
\small{$^d$ TUM-IAS and Physik Department, Technische Universit\"at M\"unchen, Munich, Germany}\\

\end{center}

\vskip 1cm

\centerline{\bf Abstract}

With the aim of linking  natural supersymmetry to flavour physics, a model is proposed based on a family symmetry $G \times U(1)$, where $G$ is a discrete nonabelian subgroup of $SU(2)$, with both F-term and  (abelian) D-term supersymmetry breaking. A good fit to the fermion masses  and mixing
is obtained with the same  $U(1)$ charges for the left- and right- handed quarks of the first two families and the right-handed bottom quark, and with zero charge for the left-handed top-bottom doublet and the the right handed top. 
The model shows an interesting indirect correlation between the correct  prediction for the $V_{ub}/V_{cb}$ ratio and large right-handed rotations in the $(s,b)$ sector, required to diagonalise the Yukawa matrix. For the squarks, one obtains  almost degenerate first two generations.
The main source of the FCNC and CP violation effects is the splitting between the first two families and the right-handed sbottom determined by the relative size of F-term and D-term supersymmetry breaking. The presence of the large right-handed rotation implies that the bounds on the masses of 
the first two families of squarks and the right handed sbottom are  in a few to a few tens  TeV range.
The picture that emerges is  light stops and left handed sbottom and much heavier other squarks.

\vskip 3mm

%%%%%%%%%%%%%%%%%%%%%%%%%%%%%%%%%%%%%%%%%%%%%%%%%%%%%%%%%%%%%%%%%%%%%%%%%%
\newpage 
\section{Introduction}

According to the hypothesis of "minimal" supersymmetry~\cite{CKN}, now termed as natural supersymmetry, the only light states in  the supersymmetric spectrum are those that have the largest
effect on the quadratic terms in the Higgs potential, and the rest is much heavier. Thus, the only light
superpartners are  the stops, the left-handed sbottom and the higgsino. Also the gluino is expected to be
moderately light, as it enters the Higgs potential at the 2-loop level. The heavy set includes, in  particular, the right handed sbottom. 
The new LHC lower bounds on the superpartner masses~\cite{SUSYATLAS,SUSYCMS},   stronger for the first two generations of squarks than for the third one, revive the interest in such a spectrum\footnote{For 
related ideas on flavor-dependent spectra see Ref.~\cite{Perez1}.} . 
Recently, several new ideas have been proposed to explain such a  spectrum \cite{Csaki} and detailed LHC analysis were performed \cite{ruderman}.

It is an obvious and  interesting theoretical question whether such hierarchical  squark  masses can be linked to the fermion mass hierarchies  in one  framework of some theory of flavour, combined with a mechanism of supersymmetry breaking.   In this paper, we show that the spectrum of minimal supersymmetry is predicted by the flavour theory
based on  family symmetry $G \times U(1)$, where $G$ is a discrete nonabelian subgroup of $SU(2)$,  with both $F$-term and $D$-term~\cite{AnomalousU1} supersymmetry  breaking. Essentially all our phenomenological considerations
are the same as in the $SU(2)_{global}$ case.
We also point out that in addition to the well known bound on  the splitting between the first two and the third family masses ~\cite{Arkani, Badziak:2012rf},  there is in general also an upper bound on the splitting between the left and right handed sbottoms,   dependent on  the value
of $\tan\beta$. It is a combined effect of the 2-loop   and 1-loop contributions to the RG equations  for the Higgs mass parameter $m_H^2$  and the scalar masses, mainly the CP-odd Higgs scalar,respectively,  that depend  on  the bottom Yukawa coupling   and are proportional to the right-handed sbottom  mass.
Thus, if the RG evolution from the high scale to the electroweak scale is involved, the "natural" spectrum is  generically  incompatible with very large values of $\tan\beta$ and with any flavour
model that implies such values, the one based on the  horizontal $U(2)$ symmetry in particular. 
The flavour model proposed in this paper   works for a large range of moderate values of $\tan\beta$.

The issues we address in this paper have a long research history. Already in the early days of supersymmetric phenomenology, a large splitting in mass between the third and the first  two families of sfermions  has been proposed as a way to reconcile the naturalness of the Higgs potential with the suppression of new potential sources  of the  FCNC and CP violation effects generated by the superpartner sector~\cite{DLK,PT,CKN}.
A very simple and attractive possibility of linking the supersymmetric spectrum to the fermion one  is offered by  flavour theories with a single horizontal  $U(1)$ gauge symmetry \cite{fn,u1}.  With both $F$-term and $D$-term~\cite{AnomalousU1} supersymmetry  breaking, one indeed predicts an inverted hierarchy of  sfermion masses  as a consequence of   the hierarchical pattern of fermion masses
\cite{Dudas,Nelson,Chankowski:2005qp}.  The  scales in the soft  masses depend on the relative magnitude of the $F$- and $D$-term breaking.  

This attractive mechanism  has been, unfortunately, facing certain phenomenological problems. In models based on a single 
$U(1)$ horizontal symmetry, the compatibility with the fermion mass spectrum implies not only the 3rd generation lighter than the first two  but also a large splitting between the first two. Thus, after the rotation from the electroweak basis to the SCKM basis, flavour off-diagonal entries  are generated and one needs $\mathcal O(1000)$ TeV   first two generation squarks to suppress FCNC effects to acceptable levels\footnote{In models with several U(1) factors the FCNC constraints can be eased by alignment, see e.g. the last references in \cite{u1} and \cite{Dudas} and Refs.~\cite{Degenerate, Perez2}.}. Such large values are not  compatible with the proper electroweak breaking at the level of 2-loop quantum corrections. The constraints on the splitting  between the left- and right-handed sbottom has neither been given a sufficient attention  so far.

In the present paper we readdress the issue of obtaining natural supersymmetry from a flavour theory and $D$-term supersymmetry breaking, inspired by the phenomenological success of the model  for  fermion masses and mixing based  on the global horizontal $U(2)$ symmetry~\cite{BDH}.
The $U(2)$ model is very predictive and describes well
the bulk of the data, with several very interesting relations between the masses and mixings.  However,
the price for good predictivity is that there  is also  some tension with the data. 
It  predicts $m_s/m_b=m_c/m_t$ (up to order
one coefficients).  Moreover, as pointed out in Ref.~\cite{RRRS}, the prediction for $|V_{ub}/V_{cb}|$ is off by a factor of 2. This tension can be traced back to the left-right symmetric structure of Yukawa matrices and the presence of texture zeros in the (1,3),(3,1) entries in the quark Yukawa matrices. As has been already proposed in Ref.~\cite{RRRS,raby}, at least one of these conditions has to be relaxed in order to achieve consistency with experiment. Since the up and  down Yukawa matrices have the same pattern, the  model naturally works only for large $\tan\beta$ values.

In our present case the left-right symmetric structure in the down sector is broken by the $U(1)$ charge assignment {\bf \footnote { With the chosen symmetry group, contrary to the $U(2)$ model, the $U(1)$ charges are free to choose, of course the same for the whole $SU(2)$ multiplets.}}.  A good fit to the fermion masses  and mixings
is obtained for several sets of the horizontal charges and in particular for  the same  $U(1)$ charges for the left- and right- handed quarks of the first two families and the right-handed bottom quark, and with zero charge for the left-handed top-bottom doublet and the the right handed top. This is the case that we investigate in detail.  The pattern of Yukawa matrices in the $SU(2)_{global}\times U(1)_{local}$ model  shows an interesting indirect correlation between  the presence of large corrections to the $V_{ub}/V_{cb}$ ratio and large  right-handed rotations in the ($s,b$)  sector required to diagonalise the Yukawas. In the limit when the Yukawa matrices become left-right symmetric, as in the $U(2)$ model, the right and left rotations become similar  and small, both of the order of magnitude of the corresponding elements of the CKM matrix, and the correction to the $V_{ub}/V_{cb}$ ratio becomes negligibly small.

In the squark sector, the $SU(2)_{global}$ symmetry ensures almost degenerate squark masses of the
first two generations.The main contribution to the FCNC and CP violation in the kaon system comes
from the splitting between the first two families and the right-handed sbottom determined by the relative size of F-term and D-term supersymmetry breaking.
An interesting conclusion  is that, in the presence of the discussed above
large right handed rotations in the ($s,b$)  sector, the experimental bounds on the right-handed sbottom and the first two families of squarks are  of the order of a few to a few tens TeV, depending on various parameters of the model and on the gluino mass. We discuss in detail that dependence, including the dependence on the structure of the K\"ahler potential. The picture that emerges is then light stops and left-handed sbottom  and much heavier other squarks. 

Finally, the $SU(2)_{global}\times U(1)_{local}$ model gives us an opportunity to investigate the mechanism of the $D$-term supersymmetry breaking in a scheme going beyond the simplest $U(1)$ case.
The presence of the additional $SU(2)$ global symmetry, to be spontaneously broken by a second flavon, imposes non-trivial constraints on the $D$-term supersymmetry breaking mechanism, particularly if a hierarchy between the $F$-term and $D$-term breaking scales is to be obtained.  Actually, we discuss all those issues with the continuous $SU(2)$  replaced by a nonabelian discrete subgroup $G$, to avoid the problem of Goldstone bosons. A discrete symmetry is also more natural from a string theory perspective. Another possibility would be to consider a weakly gauged $SU(2)$, with a gauge coupling small enough in order to suppress the non-universal D-term contribution to sfermion masses, as proposed in Ref.~\cite{BM}.

The outline of this paper is as follows. In section \ref{model} we formulate   the flavour model. In  section \ref{massesmixings} we discuss the predictions of the model for quark masses and mixings. In Section \ref{softsusy} we discuss  the soft susy breaking masses arising from both $D$ and $F$ terms.
Section \ref{FCNCconstraints} is devoted to the bounds on the squark sector obtained from the  FCNC and CP violation effects and in the leptonic sector from $\mu \to 
e \gamma$. In Section \ref{RG} we discuss the effects of the renormalization group
running starting from a high (GUT) scale and resulting constraints at low-energy. 
In section \ref{modelbuilding} we discuss the main requirements  on  models that break spontaneously both horizontal symmetries and have $D$ and $F$ term supersymmetry breaking, with a hierarchy of scales.
In section \ref{conclusions} we give our conclusions.
In two appendices we present more details of the model  and  of the calculation of the bounds on the squark masses.

%%%%%%%%%%%%%%%%%%%%%%%%%%%%%%%%%%%%%%%%%%%%%%%%%%%%%%%%%%%%%%%%%%%%%%%%
\section{The Model}
\label{model}

In the present paper we propose a flavour model based on  $G \times U(1)_{local}$ horizontal symmetry, where $G$ is a discrete nonabelian subgroup of $SU(2)_{global}$. The discrete version  (or a weakly gauged $SU(2)$) allows first of all to avoid potential Goldstone bosons after spontaneous symmetry breaking. From a UV quantum gravity perspective, whereas we can imagine continuous abelian symmetries in string theory, broken only by nonperturbative effects, there are no similar continuous nonabelian symmetries.
 On the other hand, all string theory constructions are naturally endowed with discrete symmetries. For some recent dedicated discussions on discrete
symmetries in string theory, see  e.g. \cite{heterotic} for heterotic examples and \cite{dbranes} for D-brane examples.     

%%%%%%%%%%%%%%%%%%%%%%%%%%%%%%%%%%%%%%%%%%%%%%%%%%

We have to specify the representations of the various fields under the flavour group $G\times U(1)$, where $G$ is a discrete subgroup of $SU(2)$. However, it turns out that the flavour phenomenology can be largely decoupled from the choice of $G$, so we will postpone any issues related to the global part of the flavour group to Sec.~\ref{modelbuilding}.
The simplest choice for the flavour charges is to consider $SU(5)$ invariant charges $X_{10}$ and $X_{\overline{5}}$, with Higgs uncharged. We need two flavons, an SU(2) doublet $\phi$ with charge $X_\phi$ and an $SU(2)$ singlet $\chi$ with charge $-1$. The total field content is summarized in Tab.~\ref{charges}. 
\begin{table}[h]
\centering
\begin{tabular}{c|cccccc|cc}
& $10_a$ & $10_3$ & $\fb_a$ & $\fb_3$ & $H_u$ & $H_d$ & $\phi^a$ & $\chi$ \\
\hline
$SU(2)$ & $2$ & $1$ & $2$ & $1$ & $1$ & $1$ & $\overline{2}$  & $1$ \\    
$U(1)$ & $X_{10}$ & $0$ & $\Xfb$  & $X_3$ &$0$ & $0$ & $X_\phi$ & $-1$    
\end{tabular}
\caption{Flavor group representations of the model.}
\label{charges}
\end{table}
The zero $U(1)$ charge of the 3rd generation ten-plet  takes account of the large top quark Yukawa coupling, whereas the $X_3$ is left free, to accommodate different values of $\tan\beta$.

%%%%%%%%%%%%%%%%%%%%%%%%%%%%%%%%%%%%%%%%%%%%%%%%%%%%%%%%%

The relevant part of the superpotential is given by
\begin{align}
\label{Wyuk}
W & = h_{33}^u  H_u  Q_3 U_3+ h_{23}^u Q_a U_3 H_u \frac{\phi^a}{\Lambda} \left(\frac{\chi}{\Lambda} \right)^{X_{10} + X_\phi} + h_{32}^u Q_3 U_a H_u \frac{\phi^a}{\Lambda} \left(\frac{\chi}{\Lambda} \right)^{X_{10} + X_\phi}  \nonumber \\
& + h_{12}^u H_u Q_a U_b \epsilon^{ab} \left(\frac{\chi}{\Lambda} \right)^{2 X_{10} } + h_{22}^u Q_a U_b H_u \frac{\phi^a}{\Lambda} \frac{\phi^b}{\Lambda} \left(\frac{\chi}{\Lambda} \right)^{2 X_{10} + 2 X_\phi} \nonumber \\
&+ h_{33}^d  H_d  Q_3 D_3 \left(\frac{\chi}{\Lambda} \right)^{X_3} + 
h_{23}^d Q_a D_3 H_d \frac{\phi^a}{\Lambda} \left(\frac{\chi}{\Lambda} \right)^{X_{10} + X_3 + X_\phi} + h_{32}^d Q_3 D_a H_d \frac{\phi^a}{\Lambda} \left(\frac{\chi}{\Lambda} \right)^{X_{\bar 5} + X_\phi}  \nonumber \\
&+ h_{12}^d H_d Q_a D_b \epsilon^{ab} \left(\frac{\chi}{\Lambda} \right)^{X_{10}
+ X_{\bar 5} } + h_{22}^d Q_a D_b H_d \frac{\phi^a}{\Lambda} \frac{\phi^b}{\Lambda} \left(\frac{\chi}{\Lambda} \right)^{X_{10} + X_{\bar 5}+ 2 X_\phi} \ . 
\end{align}
 We have imposed here that all exponents are non-negative
 \be
 X_{10} \ge 0, X_{3} \ge  0, X_{10}+X_{\phi} \ge  0, X_{\bar 5} + X_{\phi} \ge  0,  X_{\bar 10} + X_{\bar 5} \ge  0.
 \ee
The $h$'s are complex O(1) coefficients,  $\Lambda$ is a high flavour scale and  $a,b$ are the $SU(2)$ indices. The structure of the K\"ahler potential is discussed in section 4. In the leading order in small parameters, its details do not affect the predictions in the fermion sector. Using the flavon vevs
\begin{align}
\vev{\phi^a} & = \epsphi \Lambda \begin{pmatrix} 0 \\ 1 \end{pmatrix} \quad , & \vev{\chi} & = \epschi \Lambda \ ,
\end{align}
one can calculate masses and mixings in terms of the original parameters.

The Yukawa matrices are given by
\begin{align}
\label{yukawasu}
Y_u & = 
\begin{pmatrix}
0 & h_{12}^u \epsup & 0 \\
- h_{12}^u  \epsup  & h_{22}^u \epsu^2& h_{23}^u  \epsu \\
0 & h_{32}^u \epsu & h_{33}^u 
\end{pmatrix} \ , \\
\label{yukawasd}
Y_d & = 
\begin{pmatrix}
0 & h_{12}^d \epsup \epsd/\epsu & 0 \\
- h_{12}^d \epsup \epsd/\epsu & h_{22}^d \epsu \epsd & h_{23}^d  \epst \epsu \\
0 & h_{32}^d \epsd & h_{33}^d \epst 
\end{pmatrix} \ , 
%Y_e & = 
%\begin{pmatrix}
%0 & h_{12}^e \epsup \epsd/\epsu & 0 \\
%- h_{12}^e \epsup \epsd/\epsu & h_{22}^e \epsu \epsd & h_{23}^e  \epst \epsu \\
%0 & h_{32}^e \epsd & h_{33}^e \epst 
%\end{pmatrix} \ .
%\label{yukawase}
\end{align}

with
\begin{align}
\label{epsdef}
\epsu & \equiv \epsphi \epschi^{X_{10} + X_{\phi}}, & \epsd &  \equiv \epsphi \epschi^{\Xfb + X_{\phi}},  &
\epsup &  \equiv  \epschi^{2 X_{10}}, & \epst &  \equiv  \epschi^{X_{3}}.
\end{align}

It was noticed some time ago that in models based on Abelian gauge symmetries of the
Froggatt-Nielsen type \cite{fn} with one flavon, there are simple relations between
the mass matrices and the mixed gauged anomalies $U(1)\times G_a^2$ of the flavour $U(1)$ 
and the SM gauge group factors $G_a$ \cite{ibanezross}. These relations clearly point towards an anomaly of the Abelian flavor gauge group. Moreover, even without 
the $SU(5)$ assumption on the Abelian charges, as done here for simplicity, they also
predict the value of the weak angle $\sin^2 \theta_w = 3/8$ at the high scale \cite{ibanez}.
Those predictions rely on  Yukawa couplings generated by one flavon field and are usually
violated if there are more flavons. Our present model has three flavons from the 
$U(1)$ viewpoint, or in  $U(1) \times SU(2)$ terms, one $SU(2)$ singlet and one doublet. Interestingly enough, however, the determinants of the mass matrices depend
only on the singlet flavon field
\begin{eqnarray}
&& \det  Y_U  \ = \  (h_{12}^u)^2 h_{33}^u \ \epsilon_{\chi}^{4 X_{10}} \ , \nonumber \\
&& \det  Y_D \ = \  (h_{12}^d)^2 h_{33}^d \ \epsilon_{\chi}^{2 (X_{10} + X_{\bar 5}) + X_3} \ , \nonumber \\
&& \det  Y_L \ = \  (h_{12}^e)^2 h_{33}^e \ \epsilon_{\chi}^{2 (X_{10} + X_{\bar 5}) + X_3} \ . \label{an1}
\end{eqnarray}
By combining these mass determinants and using the values of the 
the anomaly coefficients  (in the notation of Ref.~\cite{Chankowski:2005qp}),

\begin{equation}
C_3 = C_2 = \frac{3}{5} C_1 = X_3 + 6 X_{10} + 2 \Xfb \ , \label{an2}
\end{equation}
we find the same mass-anomaly connections as in the pure $U(1)$ case with one flavon,
in particular the relation
\begin{equation}
\det  Y_U Y_D = (h_{12}^u h_{12}^d)^2 h_{33}^u h_{33}^d \epsilon_{\chi}^{C_3} \ ,  
\end{equation}
which clearly displays the need for an anomalous $U(1)$.  One crucial ingredient
in deriving these relations in our case is the vanishing of the $13$ element in the
Yukawa matrices. Therefore, models in which the $13$ element is non-vanishing, with multiple $SU(2)$ flavons with no alignment, will violate the relations above. 
The anomaly of this abelian flavor symmetry has has then to be cancelled by the Green-Schwarz mechanism~\cite{GS}. This has several well-known consequences,
among which: \\
- there is typically an induced Fayet-Iliopoulos term  close to the string scale, which has the consequence that generically the symmetry is broken at high scale.\\
- such anomalous symmetries can naturally implement supersymmetry breaking at
hierarchically small scales, combined with nonperturbative effects, which are natural in this context.
%%%%%%%%%%%%%%%%%%%%%%%%%%%%%%%%%%%%%%%%%%%%%%%%%%%%%%%%%%%%%%%%%%%%%%%%%%%%%%%%
\section{Quark masses and mixings}
\label{massesmixings}

In this section we discuss the predictions of our model for quark masses and mixings.   The model is defined at a high scale and any comparison with  experimental data is subject to renormalisation effects. Such effects give important corrections to small   CKM matrix elements due to the large top quark Yukawa coupling \cite{Olechowski:1990bh}, but not to their ratios   and to the ratios
of the first two generations Yukawa couplings to the third one. Therefore,  the scale dependence of the predictions discussed in this section is negligible. It  can only affect the fit of the small order parameters, which is anyway made with  random O(1)  coefficients.

The diagonalisation of the  Yukawa  matrices given in the previous section  by left and right rotations on the quark fields is  performed in  Appendix A. The rotation matrices, the Yukawa eigenvalues and the CKM matrix are explicitly given there. 
Here we just mention that, using the freedom of phase rotations on the quark fields, one can as usual bring the mass matrices to the diagonal form with real eigenvalues and the CKM mixing matrix can be written in the standard  form, with one physical phase. For the future calculation  of the soft sfermion masses, it is also important  that the left and right rotations depend altogether on four phases, which cannot be removed by phase redefinitions.  
\subsection{Predictions}
Before giving the results of our fit of the parameters in the Yukawa matrices to fermion masses and mixings, we discuss the predictions of the model that do  not depend on the details of that fit.
Using the results in Appendix~\ref{fermionsector} one finds, in particular,  the following relations:
\begin{align}
|V_{us}|  & \approx \sqrt{m_d/m_s} \sqrt{c_d}\,\ ,
\end{align}
\begin{align}
  |V_{ub}/V_{cb}|  & \approx  |\sqrt{m_u/m_c}+e^{i\beta}\Delta \, t_d \sqrt{c_d}|\, \ , &     |V_{td}/V_{ts}|   & \approx | \sqrt{m_d/m_s}+e^{i\beta'}\Delta \,t_d |\sqrt{c_d} \ ,       
     \label{exactrelations}
\end{align}
where   
\begin{align}
t_d &\equiv \tan \theta_d\equiv \frac{|h_{32}^d|\eps_d}{|h_{33}^d|\eps_3}\, \ ,
&\Delta & \equiv\frac{ \sqrt{m_s m_d}}{|V_{cb}| m_b} \approx0.09 \, \ ,
\end{align}
and $\beta$, $\beta'$ are phases that are given in Eq.~(\ref{Deltabeta}). 
In deriving these results we have used that $\eps_u'\ll\eps_u^2$ (as confirmed by the fit), but made no assumption yet on the relative size of $\eps_d$ and $\eps_3$.
Notice that the relations in Eq.~(\ref{exactrelations}) do not involve any unknown $\mathcal O(1)$ factors but only receive corrections of the order $\sim \eps_u^2,\eps_u'/\eps_u$. At this point, it is interesting to notice that to  obtain the same Yukawa structure as in the $U(2)$ models \cite{BDH} one needs \footnote{Note, however,  that in our case the flavon representations and vev's are different, such that scalar masses and FCNC effects in traditional $U(2)$ models
are different from the ones one would get from our models in
the particular case (\ref{epsU2}). In particular, in our case the flavon vev's can  be bigger. In what follows, when we refer to predictions of
$U(2)$ models and compare to the models in the present paper, we refer to the
original class of models in \cite{BDH}.} 
\be
\epsd = \epsu \quad , \quad \qquad \qquad  \epst = 1\, \ .
\label{epsU2}
\ee
In this case one has $\eps_d\ll\eps_3$ and it follows that $t_d\approx 0,\ c_d\approx 1$. As a consequence one obtains the stronger predictions \cite{BDH}
\begin{align}
\label{pred}
|V_{us}| & \approx \sqrt{m_d/m_s} \ , & |V_{ub}/V_{cb}| & \approx  \sqrt{m_u/m_c} \ , & |V_{td}/V_{ts}| & \approx  \sqrt{m_d/m_s} \ ,       
\end{align}
which only involve measured quantities. However our analysis shows that these predictions do not follow alone from the zero textures in the Yukawa matrices but also require $\eps_d\ll\eps_3$, or, equivalently, $Y^d_{i2}\ll Y^d_{i3}$ (see also Ref.~\cite{RRRS}).

Numerically one has, taking mass ratios at $M_Z$ and CKM elements from a global fit~\cite{Beringer:1900zz},
\begin{align}
\sqrt{m_d/m_s} & = 0.22 \pm 0.02 \ , & \sqrt{m_u/m_c} & = 0.046 \pm 0.008 \ , 
\end{align}
 \vspace{-0.9cm}
\begin{align}
|V_{us}| & = 0.2253 \pm 0.0007 \ , & |V_{ub}/ V_{cb}| & = 0.085 \pm 0.004 \ , &  |V_{td}/V_{ts}| & = 0.22 \pm 0.01 \ ,
\end{align}
so that the relations  in Eq.~(\ref{pred}) work well except the second one. Turning to the more general relations in Eq.~(\ref{exactrelations}), assuming positive interference ($\beta=0$) and comparing with the experimental values, we see that we need approximately $t_d\approx 0.5$ in order to account for the discrepancy in the second relation in Eq.~(\ref{pred}). This implies $\sqrt{c_d}\approx 0.95$, leading to $\sim 5\%$ corrections for the first and $\sim 14\%$ corrections for the third relation.  We observe an interesting correlation between
the magnitude  of the correction to the second relation in Eq.~(\ref{pred}) which goes as $\sim t_d$ and the size of the right-handed rotation matrix element $|V^d_{32}|=s_d$ (see App.~\ref{fermionsector}), which has very important implications for the discussion of the FCNC effects in Sec.~\ref{FCNCconstraints}.

In addition to the accurate relations in Eq.~(\ref{pred}), in $U(2)$ models one also gets predictions which are valid only up to combinations of $\ord{1}$ numbers
\begin{align}
\label{roughpred}
V_{cb} & \sim \sqrt{m_c/m_t} \ , & m_d/m_s & \sim m_u/m_c \ ,
\end{align}
besides the $SU(5)$ relations for the masses 
\begin{align}
m_b & \sim m_\tau \ ,  & m_s & \sim m_\mu \ , & m_d & \sim m_e \ .
\end{align}
Most of these predictions valid up  to $\ord{1}$ factors work pretty well, since
\begin{align}
 V_{cb} & \sim 0.04 \ , & \sqrt{m_c/m_t} & \sim 0.06 \ , 
 \end{align}
 \vspace{-0.9cm}
 \begin{align}
m_b/m_\tau & \sim 2 \ , & m_s/m_\mu & \sim 0.5 & m_d/m_e & \sim 5 \ .  
\end{align}
Only the second relation in Eq.~(\ref{roughpred}) requires a large numerical factor $\ord{20}$ (RG effects improve the agreement in the first relation in Eq.~(\ref{roughpred}) \cite{Olechowski:1990bh}. One possibility is to have $|h_{22}^u h^u_{33}-h^u_{23}h^u_{32}|\sim 4.5$ (see \ref{diagonalyukawasup} and \ref{diagonalyukawasdown}), which is easy to achieve with moderate deviations from $h^u_{ij}\sim 1$. This would indicate that $y_u/y_c$ is accidentally small in our parametrization while $y_d/y_s$ is at its natural value. In fact this is exactly what we need in order to explain the relative importance of the corrections to the exact relations in Eq.~(\ref{pred}).

Notice that  the $U(2)$ relations in Eq.~(\ref{epsU2}) imply that $\tan{\beta}$ is fixed to be large and another order-of-magnitude prediction is made
\be
m_u/m_t \sim m_d/m_b \ ,
\ee
which does not work as well as the other relations, being off by a factor $\ord{120}$. The two extra parameters that we have in the $SU(2) \times U(1)$ model therefore allow to accommodate also  small values of $\tan \beta$ and the different ratios $m_u/m_t$ and $m_d/m_b$. 
\subsection{Numerical Fit}
The small order parameters can be fixed by a fit to fermion masses and mixings for random $\ord{1}$ coefficients. 
In a common approach to Yukawa hierarchies arising from spontaneously broken horizontal symmetries one would typically scan over $\mathcal O(1)$ "bare" Yukawa couplings $h_{ij}$ and perform a fit for the order parameter(s) and charges. However, because of the highly constrained nature of the Yukawa couplings (which besides phases only comprise five real parameters in each sector) we have proceeded differently. In a first step, we perform a $\chi^2$ fit of the Yukawas entries $Y^d_{ij}$ and $Y^u_{ij}$ in Eqns.~(\ref{yukawasu}) and (\ref{yukawasd}) to the masses and mixings.~\footnote{We do this using real $Y_{ij}$. We do not expect the values of the charges and order parameters to depend on this simplifying assumption. We thus have 10 real parameters for 9 masses and mixings, and hence expect one flat direction, which turns out to be roughly aligned with the $Y^u_{32}$ direction.}  This results in typical values $Y^q_{ij}$ needed in any model satisfying the texture $Y^q_{13}=Y^q_{31}=Y^q_{11}=0$ and $Y^q_{12}=-Y^q_{21}$. In a second step we would like to estimate the $\epsilon_i$ of our specific model. We assume that the $h^q_{ij}$ are log normal distributed with mean $1$ and variance $\sigma=0.55$ (this particular value of $\sigma$ corresponds to the assumption that the $h_{ij}$ lie between $1/3$ and $3$ at 95\% C.L.). We thus fit the $\epsilon_i$ by minimizing 
\be
\chi^2 = \sum_{h^q_{ij}} \frac{[\log h^q_{ij}(Y^q_{ij},\epsilon_i)]^2}{2\sigma^2} 
\ ,  
\ee
where the "experimental" $Y^q_{ij}$ are the values resulting from the first step.
We obtain the result
\begin{align}
\label{fits}
\log_{10}\eps_u&=-1.5\pm 0.15 \quad , \nonumber
&\log_{10}\eps_u'&=-3.8\pm 0.35 \ , \\
\log_{10}\eps_d/t_\beta&=-2.0\pm 0.28 \quad , 
&\log_{10}\eps_3/t_\beta&=-2.0\pm 0.32 \ . 
\end{align}
Interestingly the correlations between the $\epsilon_i$ are very small. The goodness of fit is $\chi_{min}^2/(\rm d.o.f.)=0.8$. At the best fit point, all $h_{ij}$ are indeed close to unity, with the largest deviation occuring in the parameter $h^u_{22}\sim 2.8$.\footnote{
We could use the same values for the $Y^q_{ij}$ obtained in the first step and apply it to the $U(2)$ model defined by Eq.~(\ref{epsU2}). 
Taking as parameters $\eps_u$, $\eps_u'$ and $t_\beta$ we find a much worse fit with $\chi_{min}^2/(\rm d.o.f.)=3$. 
Indeed the $h_{ij}$ deviate much more from unity, with typical values $h^d_{32}\sim 10$, $h^d_{33}\sim 0.2$ etc. Put differently, with $h_{ij}$ closer to one it is impossible to reproduce the quantities $Y^u$ and $Y^d$ needed to fit the data.} These values in turn determine the original parameters according to Eq.~(\ref{epsdef})

\begin{align}
&X_{10}=\frac{1.9\pm 0.17}{|\log_{10}\epschi |} & X_{\bar 5}&=\frac{2.4\pm 0.36-\log_{10}t_\beta}{|\log_{10}\epschi|}\,\nn\\
&X_{\phi}=-\frac{0.4\pm 0.23+|\log_{10}\epsphi |}{|\log_{10}\epschi|} & X_3&=\frac{2.0\pm 0.32-\log_{10} t_\beta}{|\log_{10}\epschi|}\,.\
\label{chargesfits}
\end{align}
From the imposed relations $X_{10}+X_{\phi}\geq 0$ and $X_{\bar 5}+X_{\phi}\geq 0$, one also obtains from Eqns.~(\ref{epsdef}) and (\ref{fits}) the lower bound (at 95\% C.L.)
\begin{equation}
\epsphi >\max \,(0.016,\, 0.0028\, t_\beta) \, .
\end{equation}
Imposing that the charges are integers then gives a series of possibilities.  Here are a few examples, indicating  the acceptable range of values for the small parameters. A particularly simple possibility is for instance
\be
\label{stammcase}
\epschi\sim\epsphi\sim0.02\,\qquad X_{10}=X_5=X_3=-X_\phi=1\,\qquad t_\beta=5\,.
\ee
But the relations in Eq.~(\ref{chargesfits}) also allow for several other choices,a non exhaustive list is given in Tab.~\ref{chargeassignments}.
\begin{table}[h]
\centering
\begin{tabular}{c|ccc|cccc}
Model&	$\eps_\phi$	&$\eps_\chi$	&$\tan\beta$ 	&$X_\phi$	&$X_{10}$	&$X_5$	&$X_3$\\
\hline
A&		0.02			&0.02			&5				&$-1$		&1			&1		&1\\
B&		0.1			&0.2			&5				&$-2$		&3			&3		&2\\
B$'$&	0.1			&0.2			&20				&$-2$		&3			&2		&1\\
C&		0.2			&0.1			&50				&$-1$		&2			&1		&0		
\end{tabular}
\caption{Possible choices of parameters compatible with the fit to fermion masses and mixings.}
\label{chargeassignments}
\end{table}

Clearly, the range of values for $\epsphi=0.02-0.2$ and similarly for $\epschi$ is acceptable.
Notice that the choice $X_{\bar 5}=X_3$ is allowed for any value of $\epschi$ and in fact remains the only possibility for $\eps_\chi\lesssim 0.05$.  
This will turn out to be an important source of FCNC suppression as will be explained in detail in section \ref{FCNCconstraints}. In particular, it means that in models where $D$-term breaking dominates, the RH sbottom mass cannot be split very much from the RH down squark masses of the first two generations. For this reason we will mostly focus on the model defined by Eq.~(\ref{stammcase}), which has the additional advantage of a very small $SU(2)$ breaking order parameter, $\epsphi\sim 0.02$.

%%%%%%%%%%%%%%%%%%%%%%%%%%%%%%%%%%%%%%%%%%%%%%%%%%%%%%%%%%%%%%%%%%%%%%%%
\section{Soft SUSY breaking terms}
\label{softsusy}
The purpose of our model is to  link natural supersymmetry to  flavour physics. The structure of the soft supersymmetry breaking terms is defined at a high scale (of the order of the GUT scale) by the horizontal symmetry group   and the family charges. It is then subject to the RG evolution to the electroweak scale  and constrained by the FCNC and CP violation effects at that scale. Those constraints depend on the fits to the fermion sector and on the general structure of soft terms, which is not affected by the RG running. In this section we discuss that structure, in the next one we investigate the experimental constraints on the various low energy  scales present, once the UV structure is imposed
and  in section 6 we include the RG running effects to map the low energy picture to the UV
completion, within the considered framework.

In our scenario soft masses receive contributions both from $F$- and $D$-terms. The contributions from the $U(1)_H$  $D$-term are characterized by the scale 
\be
\tilde{m}_D = \sqrt{g_H \vev{D_H}},
\ee
while the $F$-term contributions that arise from higher-dimensional spurion operators  in  the potential suppressed by  some SUSY messenger scale $M$ have a typical size 
\be
\tilde{m}_F = F/M.
\ee
Gaugino masses, $A$-terms, the $\mu$-term and all masses of scalars uncharged under $U(1)_H$ are generated from $F$-term contributions\footnote{As will be shown explicitly, uncharged scalars can also get D-term contributions to their mass from higher-dimensional operators in the K\"ahler potential. These contributions are however suppressed by flavon vev's and for small enough vev's they are smaller than
the F-term contributions.}, while charged scalars receive an additional contribution from the $D$-term vev.  Let us first discuss the flavour structure of the $D$ terms. 
Using that $X_{10}+X_\phi\geq 0$, the K\"ahler potential for $Q$ can be given as follows:
\begin{align}
\label{kahlerQ}
 K\supset\ & |Q_a|^2+|Q_3|^2+
  z^q_{11}\, \left| Q^{\dagger a}\epsilon_{ab}\frac{\phi^b}{\Lambda}\right|^2
+ z^q_{22}\, \left|Q_b\frac{\phi^b}{\Lambda}\right|^2
+\left(z_{33}^q\frac{\phi^\dagger\phi}{\Lambda^2}
+z'^q_{33}\frac{\chi^\dagger\chi}{\Lambda^2}
\right)|Q_3|^2\nonumber\\
&+ z_{11}^{q \prime} \frac{\chi^\dagger \chi}{\Lambda^2} |Q_a|^2 +\left(
  z^q_{12}\,  Q^{\dagger a} Q_b\, \epsilon_{ac}\frac{\phi^c\phi^b}{\Lambda^2}\left(\frac{\chi^\dagger}{\Lambda}\right)^{ 2|X_\phi|} 
  + z^q_{13}\,  Q^{\dagger a}Q_3\, \epsilon_{ab}\frac{\phi^b}{\Lambda}\left(\frac{\chi^\dagger}{\Lambda}\right)^{ X_{10}-X_\phi}
  \right.
\nonumber \\
 &\left.
+  Q^{\dagger a}Q_3\, 
\left(z^q_{23}\frac{\phi^\dagger\phi}{\Lambda^2}+z'^q_{23}\frac{\chi^\dagger\chi}{\Lambda^2}\right)
\frac{\phi^\dagger_a}{\Lambda}\left(\frac{\chi^{\dagger}}{\Lambda}\right)^{ X_{10}+X_\phi}
+ h.c.  \right) + \cdots\ , 
\end{align}
where $\cdots$ denote operators which
break the $SU(2)$ symmetry down to a discrete subgroup ${\tilde D}_n$\footnote{There are several such operators. One example, for the case of the discrete subgroups ${\tilde D}_n$  (see e.g. Ref.~\cite{discrete} for the group theory), is $|Q_1|^2 |\phi^1|^2 +|Q_2|^2 |\phi^2|^2 $. We have checked that their effects  is to redefine coefficients of some of the operators in what follows, without changing our conclusions. Consequently, we will ignore writing explicitly such operators in what follows.}.
The last operator has an additional suppression factor, as otherwise it will be removed by a holomorphic field redefinition of $Q_3$. One finds the K\"ahler metric
\be
K^q_{\bar {i} j}=
\left(
\begin{array}{ccc}
1+ z^q_{11}\, \eps_\phi^2 + z_{11}^{q \prime}  \eps^2_\chi
&  z^q_{12}\, \eps_{\phi}^2\eps_{\chi}^{|2X_\phi|}
&  z^q_{13}\, \eps_\phi\eps_\chi^{X_{10}-X_\phi}\\
z^{q*}_{12}\, \eps_{\phi}^2\eps_{\chi}^{|2X_\phi|}&
1+ z^q_{22}\, \eps_\phi^2 + z_{11}^{q \prime}  \eps^2_\chi&
\left(z^q_{23}\eps_\phi^2+z'^q_{23}\eps_\chi^2\right)\eps_\phi\eps_\chi^{X_{10}+X_\phi}
\\
z^{q*}_{13}\, \eps_\phi\eps_\chi^{X_{10}-X_\phi}&
\left(z^{q*}_{23}\eps_\phi^2+z'^{q*}_{23}\eps_\chi^2\right)\eps_\phi\eps_\chi^{X_{10}+X_\phi}
&
1+\left(z^q_{33}\,\eps_\phi^2+z'^q_{33}\,\eps_\chi^2\right)
\end{array}
\right) \ . 
\label{kahlermetricQ}
\ee
The soft mass terms are found by expanding the K\"ahler potential in Eq.~(\ref{kahlerQ}) to linear order in $\langle D_H\rangle$ and going to canonical normalization using the K\"ahler metric Eq.~(\ref{kahlermetricQ}). One obtains
\begin{eqnarray}
(\tilde m^2_{q,D})_{11}&=&\left(X_{10}+z^q_{11}\, X_\phi\, \eps_{\phi}^2\right) \tilde m_D^2\nn\\
(\tilde m^2_{q,D})_{22}&=&\left(X_{10}+z^q_{22}\,X_\phi\, \eps_\phi^2\right) \tilde m_D^2\nn\\
(\tilde m^2_{q,D})_{33}&=&\left(z^q_{33}X_\phi\,\eps_\phi^2-z'^q_{33}\,\eps_\chi^2\right) \tilde m_D^2\nn\\
(\tilde m^2_{q,D})_{12}&=&2\,z^q_{12}\,X_{\phi}\,\eps_{\phi}^2\eps_{\chi}^{2|X_\phi|}\, \tilde m_D^2\nn\\
(\tilde m^2_{q,D})_{13}&=&z^q_{13}\,(X_\phi-\tfrac{1}{2}X_{10})\,\eps_\phi\eps_\chi^{X_{10}-X_\phi}\, \tilde m_D^2\nn\\
(\tilde m^2_{q,D})_{23}&=&\left[z^q_{23}\left(X_\phi-\tfrac{1}{2}X_{10}\right)\eps_\phi^2
-z'^q_{23}\left(1+\tfrac{1}{2}X_{10}\right)\eps_\chi^2\right]\eps_\phi\eps_\chi^{X_{10}+X_\phi}\, \tilde m_D^2
\label{softD10}
\end{eqnarray}
The $U$ sector works in the same way. 
For the $D$ sector we assume $|X_{\bar 5}-X_3|\leq|X_\phi|$, leading to
\begin{align}
 K\supset\ &|D_a|^2+\left|D_3\right|^2+
  z^d_{11}\, \left| D^{\dagger a}\epsilon_{ab}\frac{\phi^b}{\Lambda}\right|^2
+ z^d_{22}\,\left|D_b\frac{\phi^b}{\Lambda}\right|^2
+\left(z_{33}^d\frac{\phi^\dagger\phi}{\Lambda^2}
+z'^d_{33}\frac{\chi^\dagger\chi}{\Lambda^2}
\right)|D_3|^2
\nonumber\\
&+ z_{11}^{d \prime} \frac{\chi^\dagger \chi}{\Lambda^2} |D_a|^2 + \left(
  z^d_{12}\, D^{\dagger a} D_b\, \epsilon_{ac}\frac{\phi^c\phi^b}{\Lambda^2}\left(\frac{\chi^\dagger}{\Lambda}\right)^{ 2|X_\phi|} 
+ z^d_{23}\, D^{\dagger a} D_3\,\frac{\phi^\dagger_a}{\Lambda}\left(\frac{\chi}{\Lambda}\right)^{-X_{53}-X_\phi}
 \right.
\nonumber \\
 \label{kahlerD}
&\left. 
+\ z^d_{13}\,D^{\dagger a}D_3\,\epsilon_{ab}\frac{\phi^b}{\Lambda}\left(\frac{\chi^\dagger}{\Lambda}\right)^{ X_{53}-X_\phi}
+ h.c.\right)+ \cdots 
\end{align}
where we have defined $X_{53}=X_{\bar 5}-X_3$. Note that unlike the $Q$ and $U$ sectors, the $z_{23}$ term scales with a power of  $\chi$ rather than $\chi^\dagger$.
One finds the K\"ahler metric
\be
K^d_{\bar {i} j}=
\left(
\begin{array}{ccc}
1 + z^d_{11}\,\eps_{\phi}^2 + z^{d \prime}_{11}\,\eps_{\chi}^2&
    z^d_{12}\,\eps_{\phi}^2\eps_{\chi}^{2|X_\phi|}&
    z^d_{13}\,\eps_\phi\eps_\chi^{X_{53}-X_\phi}\\
z^{d*}_{12}\,\eps_{\phi}^2\eps_{\chi}^{2|X_\phi|}&
1 + z^d_{22}\,\eps_{\phi}^2 + z^{d \prime}_{11}\,\eps_{\chi}^2&
    z^d_{23}\,\eps_\phi \eps_\chi^{-X_{53}-X_\phi}\\
z^{d*}_{13}\,\eps_\phi\eps_\chi^{X_{53}-X_\phi}&z^{d*}_{23}\,\eps_\phi \eps_\chi^{-X_{53}-X_\phi} 
&1+\left(z^d_{33}\,\eps_\phi^2+z'^d_{33}\,\eps_\chi^2\right)
\end{array}
\right)
\ee
After going to canonical normalization one obtains the soft mass terms:
\bea
(\tilde m_{d,D}^2)_{11}&=&\left(X_{\bar 5} + z^d_{11}\, X_\phi\, \eps_{\phi}^2\right)\tilde m_D^2\,,\nn\\
(\tilde m_{d,D}^2)_{22}&=&\left(X_{\bar 5} + z^d_{22}\,X_\phi\,\eps_\phi^2\right)\tilde m_D^2\,,\nn\\
(\tilde m_{d,D}^2)_{33}&=&\left(X_3+\left(z^d_{33}\,X_\phi \eps_\phi^2-z'^d_{33}\,\eps_\chi^2\right)\right)\tilde m_D^2\,,\nn\\
(\tilde m_{d,D}^2)_{12}&=&2\,z^d_{12}\,X_\phi\,\eps_{\phi}^2\eps_{\chi}^{2|X_\phi|}\,\tilde m_D^2\,,\nn\\
(\tilde m_{d,D}^2)_{13}&=&z^d_{13}\,(X_\phi-\tfrac{1}{2}X_{53})\,\eps_\phi\eps_\chi^{ X_{53}-X_\phi}\,\tilde m_D^2\,,\nn\\
(\tilde m_{d,D}^2)_{23}&=&z^d_{23}\,(X_\phi+\tfrac{1}{2}X_{53})\,\eps_\phi \eps_\chi^{-X_{53}-X_\phi}\,\tilde m_D^2 \ . 
\label{softD5}
\eea

In principle the whole K\"ahler potential in Eq.~(\ref{kahlerQ}) and Eq.~(\ref{kahlerD}) can be multiplied by $X^\dagger X$, giving rise to $F$ term contributions to the soft masses with identical scalings with 
$\epsilon_{\phi,\chi}$, but with an additional suppression $\tilde m^2_F/\tilde m^2_D$. However, there are a few cases where the $F$ terms can be relevant. First notice that the leading $D$ term contribution vanishes for 
$(\tilde m_{I,D}^2)_{33}$, $I=q,u$, resulting in an additional suppression 
$\sim\eps_\phi^2, \eps_\chi^2$, as is explicit in the above expressions. This suppression, for small flavon vev's, is bigger than the one from the hierarchy $\tilde m_F^2\ll\tilde m_D^2$. In particular this means that the stop, the right handed stau and the left handed sbottom masses are mainly due to $F$ terms.
Second, in the particular case $X_3=X_5$, all the diagonal elements $(\tilde m_{d,D}^2)_{ii}$ are degenerate and one has to take into account the splitting induced by the $F$ terms. We then only need to consider the $F$ term contributions
\begin{align}
(\tilde m^2_{I,F})_{23}&=d^I_{23}\,\eps_\phi \eps_\chi^{X_{10}+X_\phi}\,\tilde m^2_F\, \quad ,  & I&=q,u\nonumber\\
(\tilde m^2_{I,F})_{33}&=d^I_{33}\,\tilde m^2_F\, \quad ,   & I&=q,u,d
\label{softF}
\end{align}
with other $\mathcal O(1)$ coefficients $d^I$. All other $F$ term contributions can be neglected.

Let us pause a moment and discuss the various effects of the flavour breaking terms in the K\"ahler potential. First of all, we have checked that the off-diagonal terms present in Eqs.~(\ref{softD10}) and (\ref{softD5})  give only non-leading  contributions to the exact rotation matrices diagonalising squark masses. Thus,  a very good approximation, neglecting the LR contribution, the squark mass matrices are diagonal  in the original basis.  The natural basis choice is then to perform rotations only on the fermion fields  to diagonalise Yukawa matrices, so that the flavour changing effects will appear in the quark-squark-gluino  vertices, controlled by the latter rotation angles.
 A splitting of the squark masses of the first two generations, introduced by the non-diagonal K\"ahler terms, renders the FCNCs sensitive to the large $12$ rotation angles of the quark sector, which for exactly degenerate first two generations drops out. For $\epsilon_{\phi} $ in the range 0.02 to 
0.2, as obtained from the fit in the previous section,  these effects are often subleading to the effects generated by the splitting between between
the first two and the third generation. A detailed analysis of various effects is presented in the next section.

To summarize, we now collect the relevant sfermion mass matrices in an effective parametrization, keeping only the entries that are most relevant for the SUSY spectrum and dominantly contribute to flavor changing effects.  For $\tilde{u}_L, \tilde{d}_L, \tilde{u}_R, \tilde{e}_R$ sfermions the diagonal terms also provide the main source of flavor violation 
\begin{align}
\label{m10}
\tilde{m}^2_{10} & = \mD^2 \begin{pmatrix} X_{10} & 0&  0 \\ 0 & X_{10} + c_{22}^I X_\phi \epsphi^2 & 0 \\ 0 & 0 & c_{33}^I  X_\phi \eps_\phi^2 + \tilde{c}_{33}^I  \eps_\chi^2 \end{pmatrix} + \mF^2 \begin{pmatrix} 0 & 0 & 0 \\ 0 & 0 & 0 \\
0 & 0  & d_{33}^I \end{pmatrix}_{I=q,u,e} \,, 
\end{align}
while for $\tilde{e}_L, \tilde{\nu}_L, \tilde{d}_R$ sfermions also the off-diagonal terms in Eqs.~(\ref{softD5}) can be relevant if $\eps_\phi$ is particularly large. Still, provided that 
\begin{align}
\eps_\phi^2 & < \eps_u \quad (X_3 \ne X_{\overline{5}})\, ,  &  \eps_\phi^2 & < \eps_u \tilde{m}_F^2/\tilde{m}_D^2 \quad (X_3 = X_{\overline{5}})\, , 
\end{align}
the diagonal elements also dominate flavor violating effects 
\begin{align}
\label{m5}
\tilde{m}^2_{\overline{5}} & = \mD^2  \begin{pmatrix} \Xfb & 0& 0 \\ 0 & \Xfb + c_{22}^I X_\phi \epsphi^2 & 0 \\ 0 & 0 & X_3 + c_{33}^I  X_\phi \eps_\phi^2 + \tilde{c}_{33}^I  \eps_\chi^2 \end{pmatrix} + \mF^2 \begin{pmatrix} 0 & 0 & 0 \\ 0 & 0 & 0 \\
0 & 0  & d_{33}^I \end{pmatrix}_{I=d,l}\, .
\end{align}
The structure of the $A$-terms follows the structure of the Yukawas in Eqns.~(\ref{yukawasu}), (\ref{yukawasd})   
\begin{align}
A_u & =  \mF
\begin{pmatrix}
0 & a_{12}^u \epsup & 0 \\
- a_{12}^u  \epsup  & a_{22}^u \epsu^2& a_{23}^u  \epsu \\
0 & a_{32}^u \epsu & a_{33}^u 
\end{pmatrix}, \\
A_d & =  \mF
\begin{pmatrix}
0 & a_{12}^d \epsup \epsd/\epsu & 0 \\
- a_{12}^d \epsup \epsd/\epsu & a_{22}^d \epsu \epsd & a_{23}^d  \epst \epsu \\
0 & a_{32}^d \epsd & a_{33}^d \epst 
\end{pmatrix},
\end{align}
with some complex $\ord{1}$ coefficients $a_{ij}^{u,d}$.

%%%%%%%%%%%%%%%%%%%%%%%%%%%%%%%%%%%%%%%%%%%%%%%%%%%%%%%%%%%%%%%%%%%%%%%%

\section{Flavor Constraints}
\label{FCNCconstraints}

We have shown that the hierarchy of fermion masses generates certain (inverted) hierarchy for sfermions, whose actual magnitude would depend on the relative magnitude of the scales $\tilde m_D$ and $\tilde m_F$ and on the RG renormalisation effects. In more detail, the picture that emerges is the following one:  both stop masses, the left-handed sbottom and the right-handed stau (we have neglected the LR mixing)  are  controlled by the scale $\tilde m_F$,  the first two generation squark masses  are controlled by $\tilde m_D$  (assuming $\tilde m_D>\tilde m_F$) and are necessarily heavier, whereas the mass of the right-handed sbottom and left-handed stau depend on the scale $\tilde m_D$ through the charge $X_3$.  Their dependence on
$\tilde m_F$  also   cannot be a priori neglected.  Given this general hierarchical  structure of the squark masses, in  this section we investigate if the constraints  from the FCNC and CP violation effects allow for the physical stops and the left-handed sbottom to be below or around $1$ TeV and, if taken so low, what are then the bounds on  the other squark  masses. 
Then,  in  the next section we take into account the RG evolution to map the low energy bounds into the initial conditions for  $\tilde m_D$ and $\tilde m_F$  at the high scale, where the model is defined. In the following we restrict to the most relevant observables $\epsilon_K$ and $\Gamma(\mu \to e \gamma)$ and leave a detailed analysis of the phenomenology to a future publication.
%%%%%%%%%%%%%%%%%%%%%%%%%%%%%%%%%%%%%%%%%%%%%%%%%
\subsection{Constraints from $\epsilon_K$}

The strongest constraints on the sfermion masses in this model come from $\epsilon_K$ mediated by squark-gluino exchange.  In our phenomenological analysis, we take the gluino mass in the range (1.5$-$3) TeV. Since the bounds
on  the squark masses scale inversely proportional to the gluino mass, the quoted bounds can  vary by a factor of two.
 For simplicity\footnote{For a thorough discussion of Kaon mixing in natural SUSY see e.g. Ref.~\cite{Mescia}.} we use as an estimate  the bounds on the relevant Wilson coefficients  from Ref.~\cite{UTfit}, at the scale of the soft masses. 
For the bound on the left-handed sbottom only $\Delta C_1$ is relevant (see Appendix B). Since it is proportional to the product of the left-handed rotations, which are small in the model (of the order of the corresponding  CKM matrix elements), it is not surprising that the bound on the left-handed sbottom is weak; for the gluino  mass of 1.5 TeV it is generically below
1 TeV  (it depends on the assumed values of the phases). This means that a necessary condition for natural supersymmetry is consistent in this model with the flavour data and   we take in the following stops and the left handed sbottom to be in the TeV range.  Large  rotation angles entering into the Wilson coefficients are  the right-handed rotations in the (2,3) sector, so we expect non-trivial bounds
from  ${\tilde C}_1$  and $C_4$. Although in the latter case,  one angle in the product is left-handed and small, the much stronger experimental bound on $C_4$ than on  ${\tilde C}_1$  , makes $C_4$ 
(corresponding to the LLRR amplitude in the mass insertion language) the most relevant coefficient for our discussion. 
The  imaginary part of $C_4$ is bounded by \cite{UTfit}
\begin{align}
\label{UTfitbounds}
- \frac{3.0 \times 10^{-12} }{\TeV^2} \le  {\rm Im} \, C_4 \le \frac{ 4.7 \times 10^{-12}}{\TeV^2} \ . 
\end{align}

We will now turn to the analysis  of the FCNC bounds in our model guided by the general structure of the  soft  squark masses, as given by Eqns.~(\ref{softD10}), (\ref{softD5}) and (\ref{softF}). We  calculate the supersymmetric contribution to the Wilson coefficient $C_4$  in the basis in which both quark and squark masses are diagonal. Since in model A both 1-2 and 1-3 splittings are small, we expand the masses in Eq.~(\ref{C4exact}) around the common values and use unitarity of the rotations. One obtains
\newcommand{\tf}{\tilde f_4}
\begin{align}
\Delta C_4
\label{C4general}
\ & =\ \frac{\alpha_s^2}{m_{\tilde g}^2}
\left(\hat\delta^{d,RR}_{12}\Delta^R_{31}+\tilde{\delta}^{d,RR}_{12}\Delta^R_{21}\right)  \left\{
-\frac{1}{3} \biggl[\biggr.
 \hat{\delta}^{d,LL}_{12}
	\, x_1^R 
		\,\partial_R\left(\tf(x_3^L,x_1^R) - \tf(x_1^L,x_1^R)\right) \right. \nn \\
& \left. +  \, \tilde{\delta}^{d,LL}_{12} \Delta^{L}_{21} 
	\ x_1^Lx_1^R
		\,\partial_L\partial_R\,\tf(x_1^L,x_1^R)
\biggl.\biggr] + \frac{7}{3} \biggl[\biggr. \tilde{f}_4 \to f_4 \biggl.\biggr] \right\}\ , 
\end{align}
where $x_i^{L,R} = m^2_{d^i_{L,R}}/m_{\tilde{g}}^2$ and $\Delta^A_{i1}=x^A_i/x^A_1-1$ for $A=L,R$. The loop functions
$ f_4, {\tilde f}_4$ are defined in \cite{hagelin} and  given explicitly in Eqn.~(\ref{loopfunctions2}). The details of the calculation  can be found in  App.~\ref{wilsoncoefficients}. 
The flavor suppression is encoded in the following quantities, defined as
\begin{align}
&\tilde\delta_{12}^{d,RR}\equiv (V_R^d)_{21}(V_R^d)^*_{22}\,,
&\hat\delta_{12}^{d,RR}&\equiv (V_R^d)_{31}(V_R^d)^*_{32}
=  \tilde  \delta_{12}^{d,RR} t_d^{2}\,,
\\
&\tilde\delta_{12}^{d,LL}\equiv (V_L^d)^*_{21}(V_L^d)_{22}
  \,,
&\hat\delta_{12}^{d,LL}&\equiv (V_L^d)^*_{31}(V_L^d)_{32}= \tilde\delta_{12}^{d,LL}|V_{23}^d|\left(|V_{23}^d|-\frac{m_s}{m_b}t_de^{i\alpha_d}\right)
\label{deltatildehat}
\,.
\end{align}
The product relevant for Eq.~(\ref{C4general}) is given by 
\begin{align}
\tilde{\delta}^{d,LL}_{12}  \tilde{\delta}^{d,RR}_{12} & = - \, \frac{m_d}{m_s} c_d^2  \, e^{-2 i \tilde{\alpha}_{12}} \, . 
\end{align}
The case of exactly degenerate  first two generations corresponds  to the limit $\Delta^{L,R}_{21}=0$. Due to the unitarity of the rotation matrices that diagonalise the  Yukawa matrices, those contributions are always proportional to  a product of two rotation angles $(V^d_{L,R})_{3i}$ with $i=1,2$  and nicely demonstrate the supersymmetric GIM mechanism \cite{Lalak}: they vanish in the limit of the (relevant for a given contribution) degenerate first two and the third generation squark masses. Another interesting limit is the decoupling limit  for the first two generations \cite{GNR}, where they depend only on the third generation squark  masses.  It is interesting to observe how the experimental bounds on the  Wilson coefficients result in  the bounds for squark masses as a function of the splitting between generations. 

As already mentioned in Sec.~\ref{softsusy}, the effect of the 1--2 splitting in (\ref{C4general}) is often negligible. As a rough estimate, if the splitting of the first two generation  squarks is smaller than
\be
\Delta^L_{21}\lesssim  3\log \frac{\tilde m_D^2}{\tilde m_F^2}\, |V_{23}^d|^2\,,\qquad
\Delta^R_{21}\lesssim \Delta^R_{31} t_d^2
\label{31dominance}
\ee
the corresponding terms become subleading. This is a common situation, in particular in scenario A where the smallness of the $SU(2)$ breaking results in $\Delta^{L,R}_{21}\sim \mathcal O(10^{-4})$.
However, one should keep in mind that this is not always the case, and for some parameter choices in the other scenarios in Tab.~\ref{chargeassignments} they can become the dominant source of flavour violation.

\begin{figure}
\begin{center}
\includegraphics[scale=0.7]{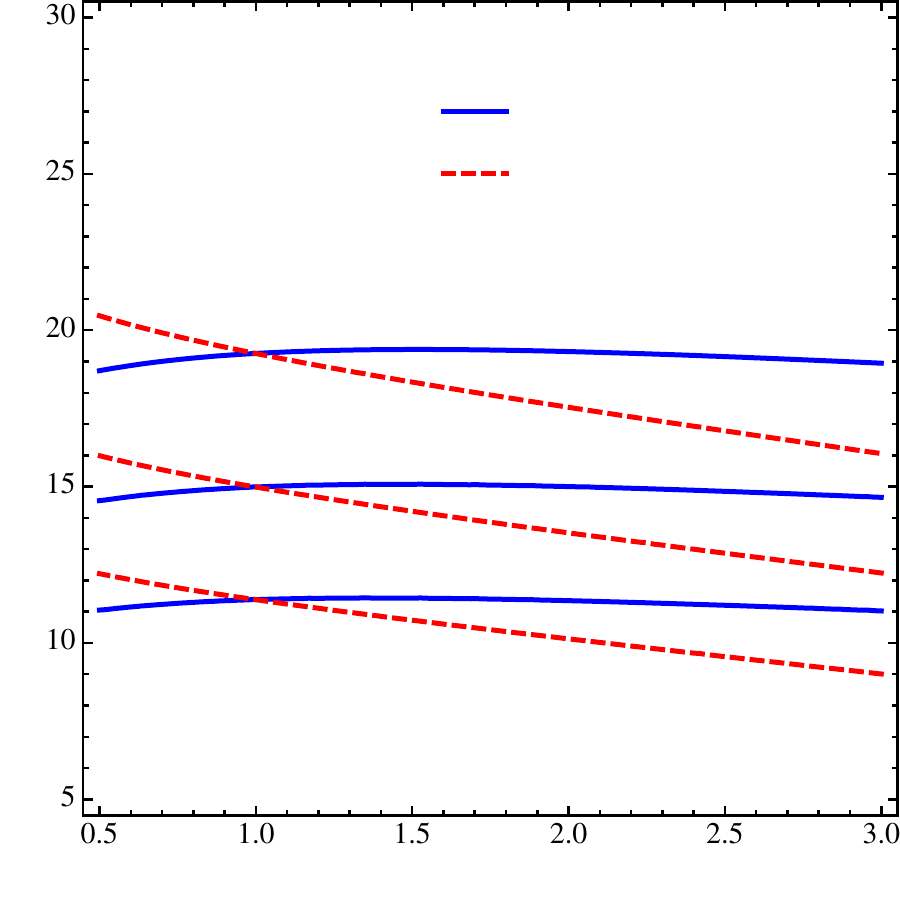}
\begin{picture}(0,0)
\put(-78,157){\footnotesize$\tilde m_{b_L}=1\ \TeV$}
\put(-78,143){\footnotesize$\tilde m_{b_L}=m_{\tilde g}$}
\put(-105,0){\footnotesize$m_{\tilde g}$ / TeV}
\put(-190,65){\rotatebox{90}{\footnotesize$\tilde m_{b_R}\approx\tilde m_{d_R}$ / TeV}}
\put(-110,28){\footnotesize$\tilde m_{d_L}=10$ TeV}
\put(-134,64){\footnotesize 1.5}
\put(-134,87){\footnotesize 2.5}
\put(-134,115){\footnotesize 4.0}
\end{picture}
\quad 
\includegraphics[scale=0.7]{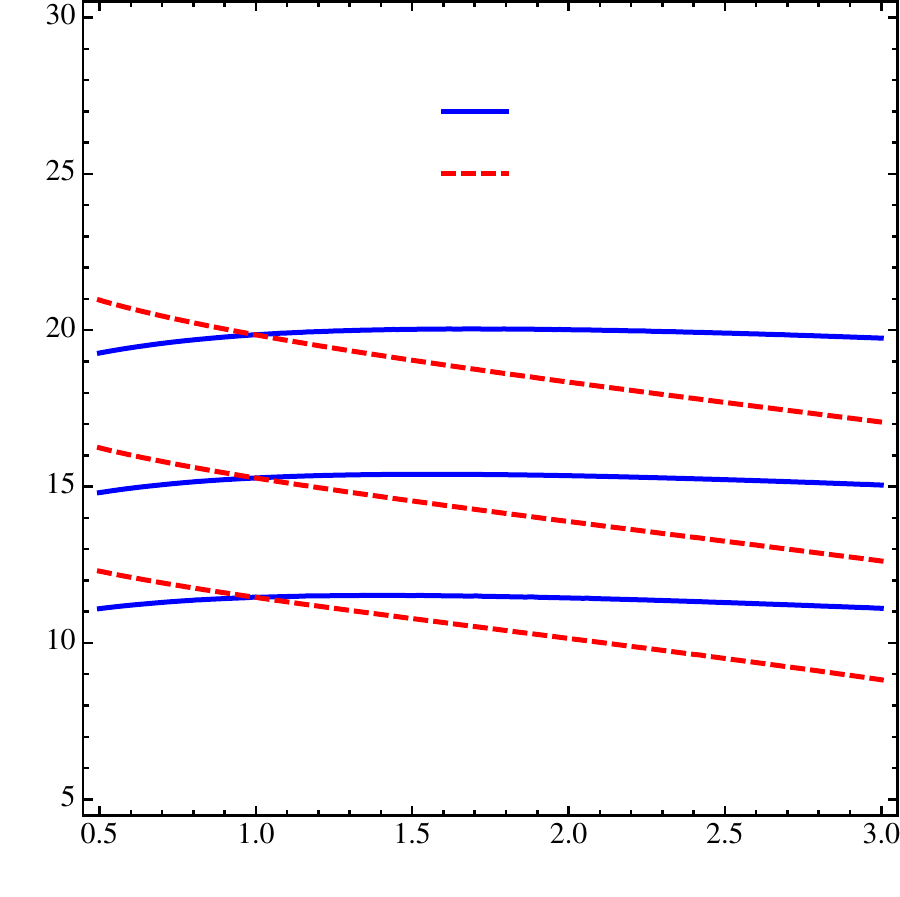}
\begin{picture}(0,0)
\put(-78,157){\footnotesize$\tilde m_{b_L}=1\ \TeV$}
\put(-78,143){\footnotesize$\tilde m_{b_L}=m_{\tilde g}$}
\put(-105,0){\footnotesize$m_{\tilde g}$ / TeV}
\put(-190,65){\rotatebox{90}{\footnotesize$\tilde m_{b_R}\approx\tilde m_{d_R}$ / TeV}}
\put(-107,28){\footnotesize$\tilde m_{d_L}=\tilde m_{d_R}$}
\put(-134,64){\footnotesize 1.5}
\put(-134,88){\footnotesize 2.5}
\put(-134,117){\footnotesize 4.0}
\end{picture}
\end{center}
\caption{Bounds on the masses of the gluino and the approximately degenerate right handed down squark sector for various choices of the parameters. The region below each line is excluded. The three lines correspond to different choices of the dominant 3-1 splitting, namely $\tilde m_{d_R}^2-\tilde m_{b_R}^2=(1.5,\ 2.5,\ 4.0\ \TeV)^2$. The remaining parameters are chosen as $|V_{23}^d|=0.04$, $\sin(\alpha_{12})=0.5$ and $s_d^2=0.2$. The decoupling of the gluino occurs outside the displayed range of the gluino mass.}
\label{boundsplot}
\end{figure}
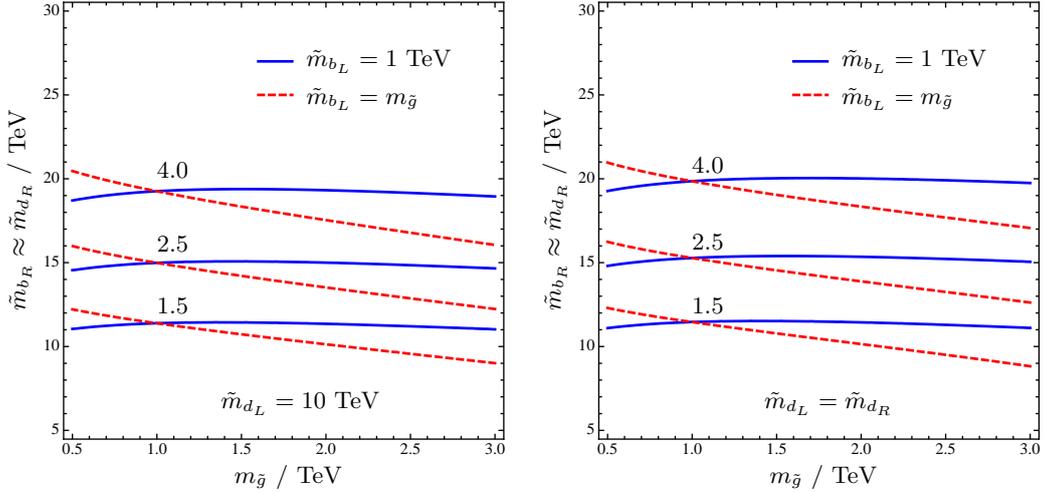
In Fig.~\ref{boundsplot} we plot the bounds arising from imposing Eqn.~(\ref{UTfitbounds}) on (\ref{C4general}) in the $m_{\tilde g}$ and $\tilde m_{b_R} \simeq \tilde m_{d_R}$ plane, for various choices of the remaining parameters.
 In the Sec.~\ref{RG} we will see that
a natural stop/gluino spectrum in the TeV range will imply a high scale vlaue of $\tilde m_F$ and hence 
a splitting $\tilde m_{d_R}^2-\tilde m_{b_R}^2$
slightly larger than $1$ TeV.  
As is clear from the various lines,
the heavy squarks (i.e. the right handed down squarks and the first two generation left handed down squarks)  should have masses around $10-20$ TeV. The bounds do not depend on the much smaller 1-2 splitting, as in model A the hierarchies in Eq.~(\ref{31dominance}) are very strong.  

If we for simplicity set 
\be
\tilde m_{b_L}=m_{\tilde g}\,,\qquad \tilde m_{d_L}=\tilde m_{d_R}
\ee
we can obtain a simple estimate for $\Delta C_4$ as a function of the remaining free masses and splittings, as well as the other parameters.
Expanding Eq.~(\ref{C4general}) for large $\tilde m_{d_R}$ one gets \footnote{Here we have neglected the term proportional to the phase $e^{i\alpha_d}$, since for typical values $V^d_{23}\sim V_{cb}\approx 0.04$ and $t_d\approx 0.5$, the second term in the last relation in Eq.(\ref{deltatildehat}) is subleading.}
\begin{align}
\label{C4b}
{\rm Im} \, \Delta C_4 & \approx  \frac{2}{3}  \alpha_s^2 \frac{m_d}{m_s}|V^d_{23}|^2 s_d^2 \,\sin 2 \tilde{\alpha}_{12} \left(\tilde m_{d_R}^2-\tilde m_{b_R}^2\right)   
\frac{\log \left(\frac{\tilde m_{d_R}}{m_{\tilde g}}\right)  + \frac{1}{4} }{(\tilde m_{d_R})^4}\nonumber \\
& \approx   1.6 \times 10^{-8} \left( \frac{|V^d_{23}|}{0.04}\right)^2 \left( \frac{s_d^2}{0.2}\right) 
\left(\frac{\sin {\alpha}_{12}}{0.5}\right) 
\left(\tilde m_{d_R}^2-\tilde m_{b_R}^2\right)  \, 
\frac{\log \left(\frac{\tilde m_{d_R}}{m_{\tilde g}}\right)  + \frac{1}{4}}{(\tilde m_{d_R})^4}
 \end{align}
where we took $\alpha_s (\mu)$ at $\mu =1.5$ TeV and used Eq.~(\ref{phase12}).

Up to now we have been working in model A. For the other models consistent with the quark masses and mixings, summarized in Tab.~\ref{chargeassignments}, the bounds are much stronger. The reason is that the 3--1 splitting is not set by $\tilde m_F^2$ but instead by $\tilde m_D^2$ (as $X_3\neq X_5$). Moreover, the 2--1 degeneracy is also less exact due to the much larger $SU(2)$ breaking, $\epsphi\sim 0.1-0.2$. Barring some fine tuning of parameters, we then need to go to scales at least of the order of $\tilde m_{b_R}\sim\mathcal O(100\ \TeV)$. Given that these models also require generally large $\tan\beta$, such hierarchies can easily run into problems of RG induced tachyonic masses for squarks and sleptons. We will quantify this statement in the next section.
%%%%%%%%%%%%%%%%%%%%%%%%%%%%%%%%%%%%%%%%%%%%%%%%%%%%%%%%%%%%%
\subsection{Constraints from $\mu \to e \gamma$}

Another important effect in these class of models is the contribution to BR($\mu \to e \gamma$) through the exchange of charginos and sneutrinos. Although all three sneutrinos are at the heavy scale $\tilde{m}_D$, this contribution is enhanced by large LH mixing angles in the charged lepton sector (that are of the same order as the RH mixing angles in the down sector). Therefore it dominates over the contribution from neutralinos and right-handed staus in the loop, which is strongly suppressed by 
CKM-like mixing angles, and typically also over the contribution from neutralinos and left-handed staus, which has the same flavor and mass suppression, but smaller numerical coefficients. 

\begin{figure}
\begin{minipage}[b]{0.45\linewidth}
\centering 
\includegraphics[scale=0.25]{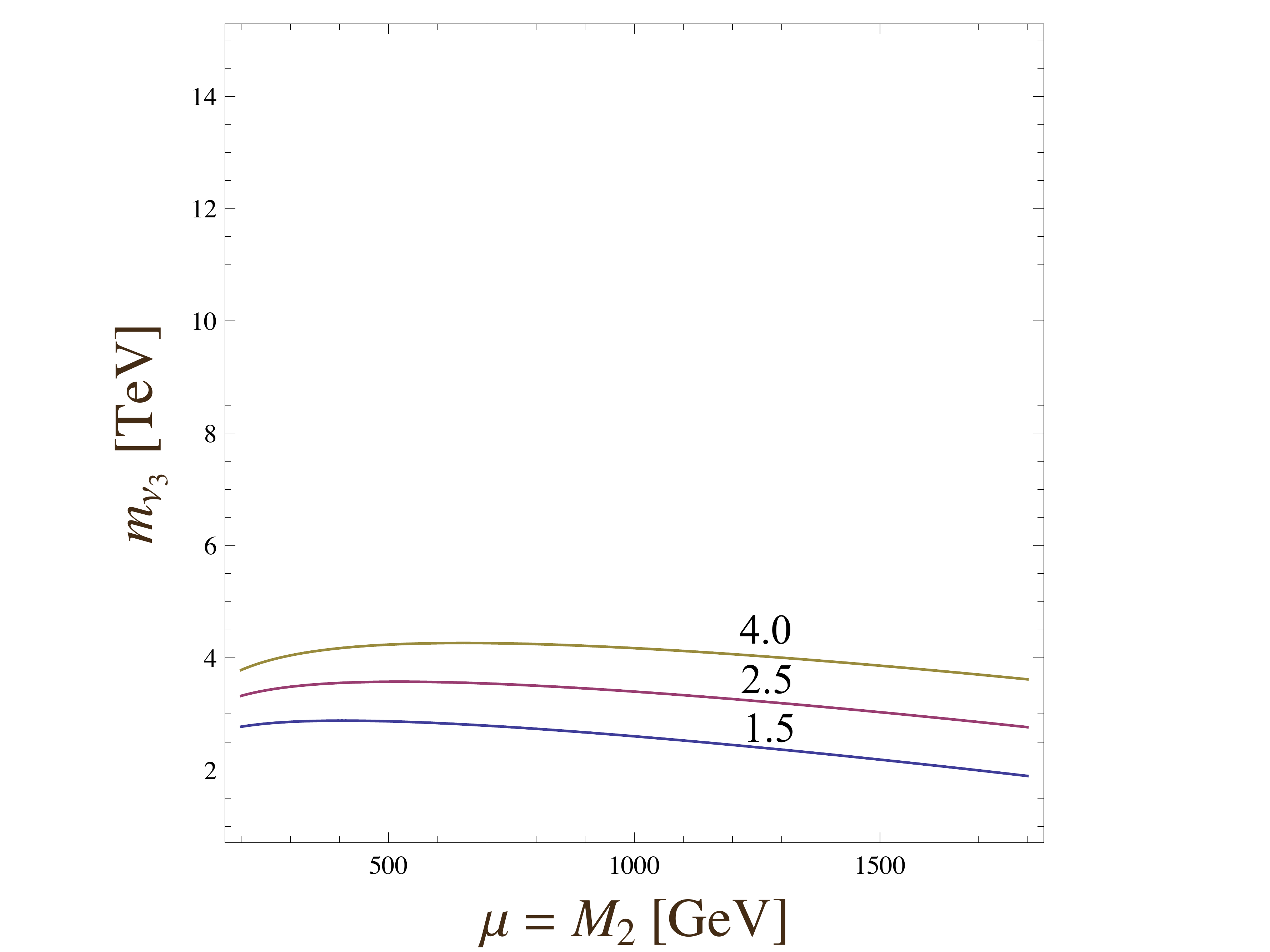} 
\end{minipage} 
\quad
\begin{minipage}[b]{0.45\linewidth}
\centering 
\includegraphics[scale=0.25]{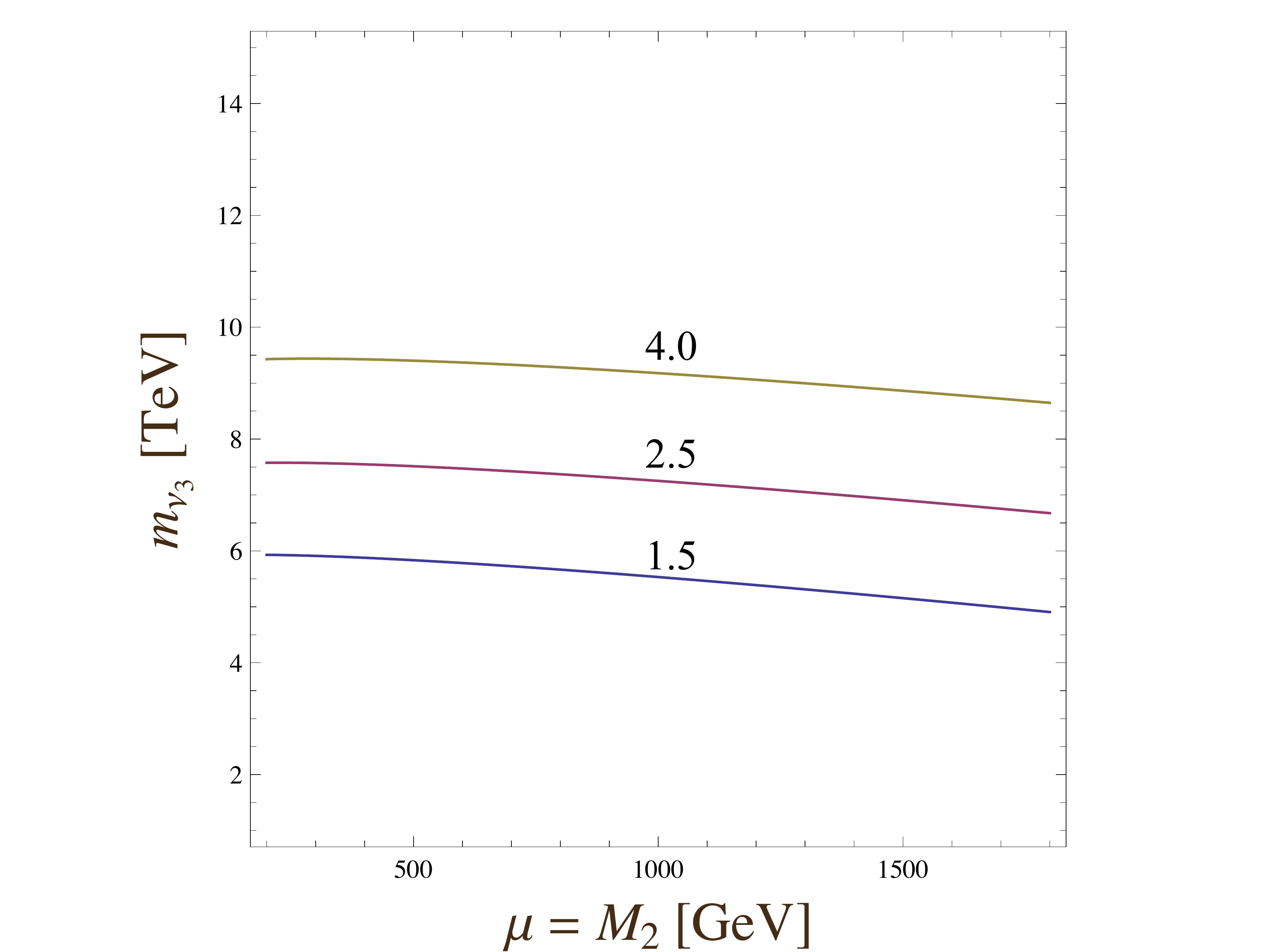} 
\end{minipage} 
\caption{Allowed region from BR($\mu \to e \gamma$) constraint in the $m_{\tilde{\nu}_3}$/$\mu $ plane for $\mu = M_2$. The region below each line is excluded. The three lines correspond to different choices of the 3--1 splitting, $\Delta m^2 =(1.5,\ 2.5,\ 4.0\ \TeV)^2$, and $\tan \beta = 5$ (left panel), $\tan \beta = 50$ (right panel). }
\label{snu3bounds}
\end{figure}

For the calculation of the decay rate we use the general results of Ref.~\cite{Hisano}. We then expand the chargino mixing matrices in leading order in $m_W/M_i$ and $m_W/\mu$, neglect LR mixing and use approximate degeneracy of the first two sneutrino generations to carry out the flavor summation. The result for the decay rate is
\be
\Gamma(\mu \to e \gamma) \approx \frac{\alpha}{4} m_\mu^5 |A_2^{(c)R}|^2 \, ,
\ee
with 
\be
A_2^{(c)R} \approx - \frac{\alpha}{8 \pi s_W^2} (Z_\nu)_{31} (Z_\nu)^*_{32} \left[ \frac{f(\mu, M_2, \tan \beta, \tilde{m}^2_{\nu_3})}{\tilde{m}^2_{\nu_3}} - \frac{f(\mu, M_2, \tan \beta, \tilde{m}^2_{\nu_1})}{\tilde{m}^2_{\nu_1}} \right] \, ,
\ee 
and the loop function can be found at the end of Appendix B. Here the matrix $Z_\nu$ diagonalizes the sneutrino mass matrix in the super-CKM basis and is therefore given by the LH charged lepton rotation matrix in the limit where sneutrino masses are diagonal in the original basis (i.e. the diagonal D-term contribution dominates), $Z_\nu \approx V^e_L$. Then using the approximate SU(5) relation $V^e_L \approx V^d_R$ one finds for the relevant flavor transitions
\be
| (Z_\nu)_{31} (Z_\nu)_{32} | \approx  |V^d_{32}|^2 |V^d_{12}|/c_d \approx  \sqrt{\frac{m_e}{m_\mu}} \frac{s_d^2}{\sqrt{c_d}} \approx 0.15 \, ,
\ee
where we used $c_d \approx 0.9$. Similarly the loop function can be estimated very roughly by neglecting the weak dependence on $M_2$ for $M_2 \approx \mu \ll \tilde{m}_{\nu_3}$. Introducing the parametrization $\tilde{m}_{\nu_1}^2 = \tilde{m}_{\nu_3}^2 + \Delta m^2$, the 3--1 splitting is approximately the same as in the right-handed sdown sector, $\Delta m^2 \approx \tilde{m}_{d_R}^2 - \tilde{m}_{b_R}^2$, so that the loop function can be expanded for small splittings $\Delta m^2 \ll \tilde{m}_{\nu_3}$ corresponding to $X_3 = X_5$. Within these rough approximations the branching ratio is given by  
\be
BR(\mu \to e \gamma) \approx 1.1 \times 10^{-11}  \left( 11 + 3 \tan \beta \right)^2 \left( \frac{\Delta m^2}{\tilde{m}^2_{\nu_3}}\right)^2  \left( \frac{\TeV}{\tilde{m}_{\nu_3}}\right)^4 \, , 
\ee
to be compared with the bound provided by the MEG Collaboration~\cite{MEG}
\be
BR(\mu \to e \gamma) < 5.7 \times 10^{-13}.
\ee
For the exact loop function, the bounds  in the $M_2-\tilde{m}_{\nu_3}$ plane for $\mu = M_2$ and different values of $\Delta m^2$ are shown in Fig.\ref{snu3bounds}. 

%%%%%%%%%%%%%%%%%%%%%%%%%%%%%%%%%%%%%%%%%%%%%%%%%%%%%%%%
 \section[]{Mapping to the high scale\footnote{This section is written in collaboration with M.Badziak.}}
\label{RG}

A brief summary of our results from the previous sections is as follows. We have defined a flavour theory at a high (of the order of the GUT)  scale  by the underlying  flavour symmetry. The horizontal charges have been determined by a fit to the quark masses and mixing. This procedure does not introduce any relevant scale dependence. We have chosen Model A in Table 2 with $X_{10}=X_5=X_3 =1$ and 
$\eps_{\phi} \approx \eps_{\chi} \approx 0.02$ as our reference model. This model gives the structure of the squark masses at the high scale in terms of the two unknown mass scales  ${\tilde m}_D$ and ${\tilde m}_F$. The stops, the left-handed sbottom and the right-handed stau  masses,  as well as the  gaugino masses depend only on ${\tilde m}_F$.
The ${\tilde m}_D$ contributions to the remaining  sfermion masses are universal (in Model A) and the first two generations can be split from the right-handed sbottom/left-handed stau by a contribution of order ${\tilde m}_F$.  Apart from quantitative aspects of the RG evolution from the high to the electroweak scale, the basic structure  at the low scale remains: the stops, the left-handed sbottom and the gluino physical masses continue to be of the same order of magnitude; the first two generations and right-handed sbottom/left-handed stau  are heavier,  with a splitting of the latter from the first two (degenerate) generations of the order of the lower mass scale.

%Clearly, independently of the quantitative aspects of the RG evolution from the high to the electroweak scale, the basic structure at the latter scale is the same: the stops, the left-handed sbottom and the gluino physical masses continue to be of the same order of magnitude; the first two generations can be heavier, with the splitting of the order of the former  masses from the right-handed physical sbottom mass.

Within that structure, we have investigated the constraints from the CP violation in the Kaon system on the above two mass scales set by the physical squark masses, for  gluino mass in the range $1-3$ TeV. We have found that the bounds on the stop and the left-handed sbottom mass  is below $1$ TeV.  Taking those masses to be indeed in the $1$ TeV range, as suggested by naturalness, we have then found
that the lower bound on the physical mass of the heavy set is in the range of $10-20$ TeV. Thus, our model accommodates -- in fact hints to -- the desired spectrum of minimal supersymmetry by linking it to a flavour theory, with a sizable splitting by a factor $10-20$ between the physical masses of the light and heavy sets of superpartners.

The questions we are now facing are : i) what are the initial conditions in terms of the high scale parameters ${\tilde m}_D$ and ${\tilde m}_F$ for such a hierarchical  spectrum  and ii) can they be consistent with the proper electroweak breaking and the absence of any tachyonic states?  It is well known that in models defined at a high scale,  the initial   splitting of the first two families of squarks (let their mass be ${\tilde m}_{1,2}$) from the  third one (with mass ${\tilde m}_3$)  cannot be   arbitrarily large (see Refs.~\cite{Arkani,Badziak:2012rf}). This is because of the 2-loop sensitivity  to the heavy masses in the RG equations for the Higgs mass parameter $m^2_{H_u}$ and for the stop masses. The numerical solutions to the 2-loop RG equations in terms of the intial values of the SO(10) symmetric soft mass parameters at the GUT scale read (obtained  in Ref.~\cite{Badziak:2012rf} 
with the SOFTSUSY code ~\cite{Allanach:2001kg}):

\begin{align}
m^2_{H_u}& \approx -1.3M^2_{1/2}-0.1A^2_0+0.35M_{1/2}A_0+0.01{\tilde m}_3^2+0.006{\tilde m}_{1,2}^2 \, \nonumber \\
{\tilde m}_{q_3}^2 & \approx 3.1M^2_{1/2}-0.04A^2_0+0.1M_{1/2}A_0+0.65{\tilde m}_3^2-0.03{\tilde m}_{1,2}^2 \, \nonumber  \\
{\tilde m}_{t_R}^2 & \approx 2.3M^2_{1/2}-0.07A^2_0+0.2M_{1/2}A_0+0.35{\tilde m}_3^2-0.02{\tilde m}_{1,2}^2 \,\nonumber \\
\tilde{m}^2_{\tau_R} & \approx 0.13M^2_{1/2}-0.055A^2_0+0.035M_{1/2}A_0+0.98{\tilde m}_3^2-0.002{\tilde m}_{1,2}^2. 
\end{align}
The coefficients in the above equations have been computed at the scale $Q=1.5$ TeV, $\tan\beta=10$, $M_{1/2}=700$ GeV, $A_0=-3 \, \TeV$, ${\tilde m}_3=3 \, \TeV$ and ${\tilde m}_{1,2}=10 \, \TeV$. Their precise values depend slightly on that choice but the qualitative features of the solutions remain unchanged. To a very  good approximation, the coefficients in Eq.~(6.1) are as obtained in the limit of vanishing bottom Yukawa coupling (the effects due to the Yukawa couplings of the first two generations are totally negligible anyway). Its effect is small since it is multiplied by small masses of the light set ( and introduces  only a very mild $\tan\beta$ dependence).

The positive contribution of the $\tilde m_{1,2}$ to the running of $m_{H_u}^2$ may destroy electroweak breaking whereas the negative contribution of  $\tilde m_{1,2}$  to the  running of the stops and left-handed sbottom masses  can  drive them tachyonic. For a fixed value of $\tilde m_3$, the synergy of both effects leads to an upper bound on the mass $\tilde m_{1,2}$
and for a fixed large $\tilde m_{1,2}$, a lower and upper bound on $\tilde m_3$, constraining the allowed parameter space.
Fixing the physical stop masses to be around
$1$ TeV, one obtains an upper bound on the hierarchy of the physical masses (since the first two generation masses run very weakly).
This bound is rising with the gluino mass, as the gluino contribution eases both  effects of $\tilde{m}_{1,2}$.

In our present case the initial spectrum is only $SU(5)$ symmetric and also the right-handed sbottom and  the left-handed stau are heavy. 
%In our present case, there is also a  splitting between the left- and right-handed sbottom and  a %similar effect for staus (with left and right interchanged), since in our model the initial spectrum  is %$SU(5)$ symmetric.
This splitting introduces  important corrections  to the solutions in Eq.~(6.1) that lead to a $\tan\beta$ dependence in both effects above, strengthening them for larger values of $\tan\beta$. This information is encoded already in the 1-loop RG equations, which include
the terms $Y^2_b{\tilde m}^2_{b_R}$  and $Y^2_\tau{\tilde m}^2_{\tau_L}$ :

\begin{eqnarray}
&& \frac{d}{dt} {\tilde m}^2_{q_3}=\frac{16}{3}g_3^2M^2_3 + \cdots - Y^2_b{\tilde m}^2_{b_R} \, \nonumber \\
&& \frac{d}{dt} {\tilde m}^2_{\tau_R}= \frac{12}{5} g^2_1 M^2_1 + \cdots - 2Y^2_\tau{\tilde m}^2_{\tau_L}
\, 
\end{eqnarray} 
and
\begin{equation}
\frac{d}{dt} m^2_{H_u}=3g_2^2M^2_2+ \frac{3}{5} g^2_1 M^2_1-3Y^2_t({\tilde m}^2_{q_3}+{\tilde m}^2_{t_R}+m^2_{H_u}+A^2_t) \, 
\end{equation}
where $t=\frac{1}{8\pi^2}\log\frac{M_{GUT}}{Q}$ and  $M_i$ are gaugino masses. 

Adopting the notation of this paper, we shall denote the light masses at high scale  by ${\tilde m}_F$ and the heavy ones by ${\tilde m}_D$.
The convolution of the  equation for $ m^2_{H_u}$ with the equation for ${\tilde m}^2_{q_3}$ gives the net positive contribution\footnote{This and the following numbers  are an approximate parametrization of the numerical results obtained with the SOFTSUSY code.} to the former, approximately  $+0.005(\tan\beta/30)^2{\tilde m}_D^2$, whereas
the contributions  to ${\tilde m}_{t_L}^2$, ${\tilde m}_{t_R}^2$  and ${\tilde m}_{\tau_R}^2$   are approximately
$ -0.03(\tan\beta/30)^2{\tilde m}_D^2 $, $+0.003(\tan\beta/30)^2{\tilde m}_D^2$   and $-0.055 (\tan\beta/30)^2{\tilde m}_D^2$  , respectively. These are additional terms that have to be added to
the solutions in Eq.~(6.1), with the obvious identification ${\tilde m}_{1,2}^2={\tilde m}_D^2$ and ${\tilde m}_3^2={\tilde m}_F^2$.~\footnote{To be precise, the shift of right-handed sbottom and left-handed stau from the light to the heavy set leads also to a very mild change in  the coefficients of the terms already present in Eq.~(6.1).}

Finally, a very important constraint on the high energy  parameter space  comes  from the experimental bounds on  the CP-odd Higgs scalar with mass $m_A$. Using the weak scale condition for the correct electroweak symmetry breaking (at intermediate and large $\tan\beta$ values)
\begin{equation}
m_{H_u}^2\approx -\mu^2,
\end{equation}
one has the following relation at  the electroweak scale for $m_A$:
\begin{equation}
m_A^2=m_{H_u}^2+m_{H_d}^2  + 2\mu^2\approx m_{H_d}^2-m_{H_u}^2.
\end{equation}
Thus, one can express the CP-odd scalar mass in terms of the high energy parameters by solving
the RG equation for $(m_{H_d}^2-m_{H_u}^2)$

\begin{equation}
\frac{d}{dt}(m_{H_d}^2-m_{H_u}^2)=  3Y^2_t{\tilde m}^2_{q_3} + \cdots - 3Y^2_b{\tilde m}^2_{b_R}-Y^2_{\tau}{\tilde m}^2_{\tau_L}.
\end{equation}
The net  ${\tilde m}_D^2$ dependent contribution to $m_A^2$ is approximately
$-0.12(\tan\beta/30)^2{\tilde m}_D^2 $, so it is
pushed by the terms proportional to ${\tilde m}_D^2$ into the "tachyonic" region more effectively  than the sfermions. The experimental bounds on $m_A$ as a function of $\tan\beta$ therefore put an important lower bound on ${\tilde m} _F$ (that controls the positive contributions to $m_A$), for a large fixed value of ${\tilde m}_D$.

The synergy of the  required proper electroweak breaking  and the experimental limits on $m_A$ leads now to much stronger bounds  in the space (${\tilde m}_D, {\tilde m}_F)$ as a function of $\tan\beta$.    In consequence, one gets strong bounds on the hierarchy of the physical masses of the two sets of superpartners. Turning it around, the natural physical spectrum of our model, with stops and the left-handed sbottom around  $1$ TeV and with the other squarks a factor $10-20$ heavier, cannot be realized for arbitrarily large values of $\tan\beta$. 

\begin{figure}
\begin{center}
\includegraphics[scale=0.8]{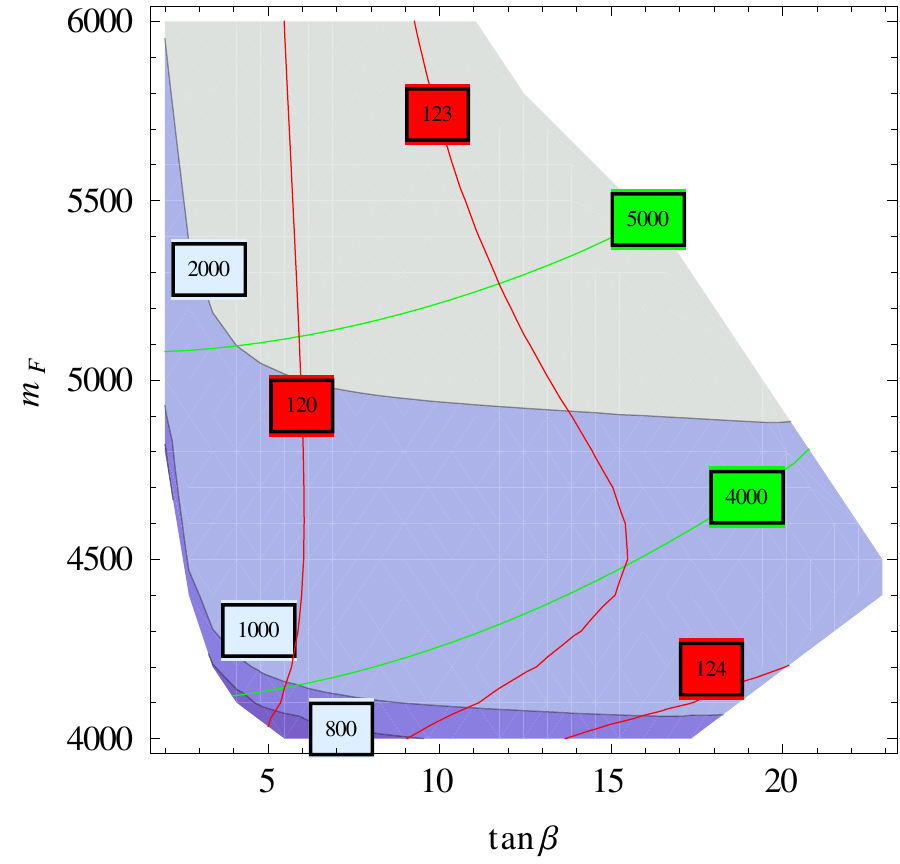}
\quad 
\includegraphics[scale=0.8]{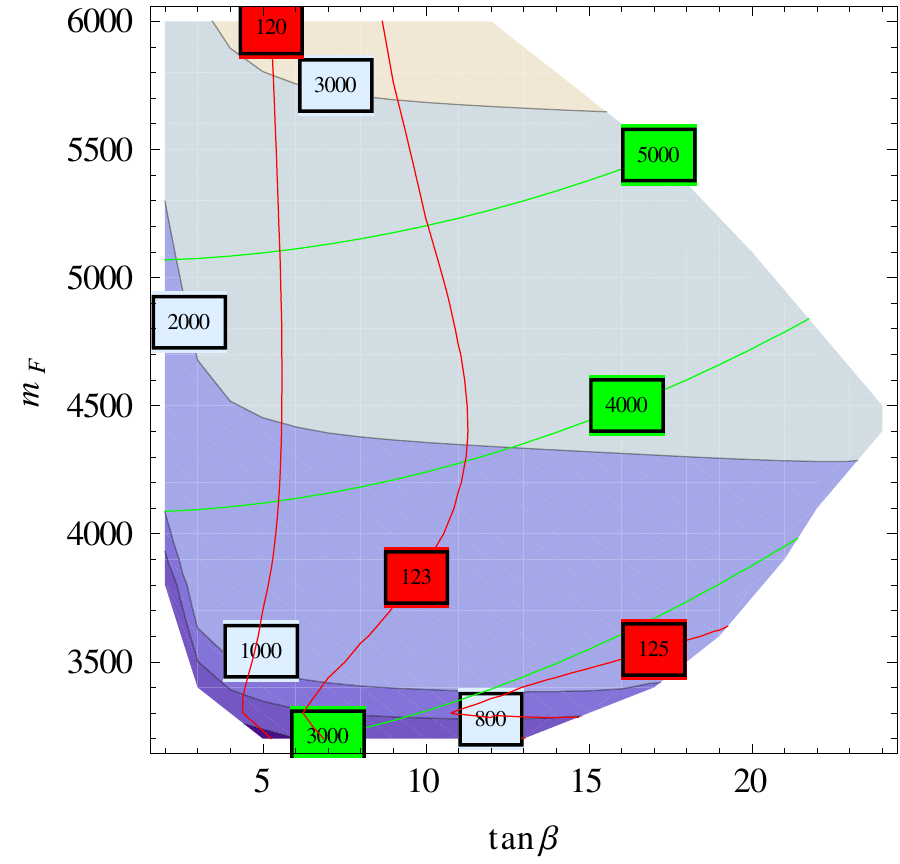}
\end{center}
\caption{Parameter region in the $\tilde{m}_F/ \tan \beta$ plane for fixed $\tilde{m}_D = 15 \, \TeV$ and $M_{1/2} = 0.6 \, \TeV, A_0 = -3.0 \, \TeV$ (left panel) and $M_{1/2} = 1.0 \, \TeV,  A_0 = -2.0 \, \TeV$ (right panel). The contour lines correspond to the masses of $\tilde{t}_1$ (blue), $\tilde{\tau}_1$ (green) and $h^0$ (red).}
\label{Marcinplots}
\end{figure}

Those qualitative considerations are nicely illustrated by the two
plots in Fig.~\ref{Marcinplots}.  In both plots the value of ${\tilde m}_D$ is fixed to 15 TeV, and A-terms are chosen as $A_0 = - 3 \, \TeV$ (left plot) and $A_0 = - 2 \, \TeV$ (right plot). The initial gluino mass is fixed to 0.6 (1.0) TeV in the left (right) plot respectively, corresponding to approximately 1.5 and 2.5 TeV physical gluino masses. Since the first two generation and the R-handed sbottom masses run  very weakly,  their physical masses   remain around 15 TeV. The blue, green and red colours describe the lighter stop, the lighter stau and the Higgs  mass values, respectively. In both plots we see the  $\tan\beta$ dependent upper bound on ${\tilde m}_F$ coming from the requirement of proper electroweak breaking. The lower bound on ${\tilde m}_F$ does not depend on $\tan\beta$ for certain range of its values  for which it is due to the the 2-loop stop "tachyonic" constraint that is independent of the left- and right- handed sbottom splitting.
For large enough $\tan\beta$ the 1-loop splitting effects take over, and the lower bound on ${\tilde m}_F$  increases with $\tan\beta$
as a result from the experimental limit on $m_A$. The intersection of the upper and lower bound on $\tilde{m}_F$ then determines the allowed upper value of $\tan\beta$ . There is also an exclusion region for very low $\tan\beta$ which depends on the details of the RG equations.

%Note however that the values
%of $\tan\beta$  with stops around 1 TeV are  below 18 ( 17) for $M_{1/2}=0.6(1)$ TeV. It is intuitively
%clear that the heavier the gluino the stronger the bound for $\tan\beta$  admitting  a light stop %because for heavier gluinos it is the limit on $m_A$ that puts the lower bound on ${\tilde m}_F$.  
%It is also clear that for a heavier gluino,  the absolute (for the used in the plots values of the other %parameters) bound on the allowed value of $\tan\beta$ corresponds to much heavier stops because it %requires  larger values of ${\tilde m}_F$ for keeping $m_A$ above the experimental limit. }

Finally, it is interesting  to see that, for the parameter values of the plots, the lighter  right-handed stau mass remain close to its initial value ${\tilde m}_F$
because for  ${\tilde m}_{\tau_R}^2$ in Eq.~(6.1) the coefficient of  ${\tilde m}_F^2$ is close to 1 and the negative 2-loop effect from $\tilde{m}^2_{1,2}$ is an order of magnitude smaller than for stops.  Although the $\tan\beta$ dependent effects summarized below (6.3) are similar, as the net result the  right-handed stau mass runs very little. Therefore the constraints from LFV are typically satisfied, as the stau is significantly heavier than the stop. For larger values of the gluino mass, however, the stau mass gets closer to the stop mass since the latter gets a larger contribution  from gluino renormalisation.

Summarizing,  the minimal supersymmetry  spectrum of our model fits nicely a two  mass scale initial
conditions with a moderate hierarchy $\frac{{\tilde m}_D}{{\tilde m}_F}\approx (3-5)$. However, the  range  of  $\tan\beta$   for which such a spectrum can be obtained is limited  to small and intermediate values  as a result from proper EWSB.  This is an interesting constraint on minimal natural supersymmetry, with only stops and the left-handed sbottom light and the rest of squarks heavy,  almost totally model independent. 
%Flavour models requiring very large values of $\tan\beta$ are in conflict with a large splitting between left-  and right-handed sbottom masses, but of course natural supersymmetry does not have to be the minimal one.
In the framework of our $SU(2)\times U(1)$ flavour model, the constraints discussed in this section strongly point to an almost unique choice of  universal U(1) charges $X_{10}=X_5=X_3$, since larger mass splittings would significantly rise the lower bound on the heavy set masses (see section 5), in conflict with the results of the present section.

%%%%%%%%%%%%%%%%%%%%%%%%%%%%%%%%%%%%%%%%%%%%%%%%%%%%%%%%%%%%%%%%%%%%%%%%%%%%
\section{Model building requirements}
\label{modelbuilding}
All our discussion of fermion and superpartner masses in the previous sections are compatible with  a flavor symmetry based on $U(1) \times SU(2)$, with broken supersymmetry, and where $U(1)$ is gauged in order to provide large $D$-terms $D> (F/M)^2$ by a factor of $3-5$, whereas $SU(2)$ is a global symmetry. The model would link then flavour symmetry with natural supersymmetry in a simple and economical way. It is clearly an interesting  challenge to put these ideas on a more firm theoretical footing.

From a theoretical viewpoint  a continuous $SU(2)$ is problematic, since after spontaneous symmetry breaking it leads to massless goldstone bosons. In a string theory setup, there are no obvious ways to obtain continuous non-abelian symmetries, whereas discrete nonabelian symmetries are typically present and related to the geometry of the internal space \cite{heterotic} and/or of the fluxes needed for generating chirality in realistic models \cite{dbranes}. Another potential possibility would be to consider a gauged $SU(2)$ with a gauge coupling small enough in order to prevent non-universal D-term contributions to soft masses. Both options are worth further exploration towards realistic UV completions. Here we limit ourselves to some remarks on discrete symmetries. 

The simplest discrete symmetries that do the job seem to be
the  groups ${\tilde D}_n$, also called $D'_n$ in the literature. However, most renormalizable operators  preserve $SU(2)$ and typically at renormalizable  tree-level  there are still
massless goldstones.  Most lower higher-dimensional operators that
break $SU(2) \to {\tilde D}_n$ preserve actually the continuous subgroup $U(1) \in SU(2)$ 
and as such, there is still one massless goldstone. Finding higher-dimensional operators that break also this Abelian subgroup and give mass to the goldstone is therefore necessary. 
         
Moreover, the model should  induce the structure of vevs that we need: large scalar vevs for the flavons and alignment in the flavor space in order to provide successful fermion, scalar
mass matrices, $A$-terms and gaugino masses.

To summarize, the minimal setup that seems viable is based on supersymmetric models with flavor symmetries $U(1)_X \times {\tilde D}_n$,  satisfying the following requirements: 
\begin{itemize}
\item Supersymmetry is broken with D and F-terms, such that  $D > (F/M)^2$ (by a factor of $3-5$). 
\item The dynamics of the model generate appropriate vev's and vacuum alignment. 
\item  There are higher-dimensional operators that break explicitly 
$SU(2) \to {\tilde D}_n $, in particular breaking the subgroup $U(1) \in SU(2)$.
\end{itemize}
A complete theoretical framework should be a string theory with an anomalous flavor-dependent $U(1)$ gauged symmetry and non-abelian
discrete subgroup of $SU(2)$. The vacuum structure is typically tied also to
moduli stabilization, due to the anomalous nature of $U(1)$ and the field-dependence
of the Fayet-Iliopoulos terms in string theory. Whereas a complete string theory model along these lines is beyond the goals of our paper, it is worth mentioning 
here that the main ingredients of our construction are present in  most current realistic string theory constructions.     
 %%%%%%%%%%%%%%%%%%%%%%%%%%%%%%%%%%%%%%%%%%%%%%%%%%%%%%%%%%%%%%%%%%%%%%%%%%
\section{Conclusions}

The first round of the LHC experiments suggests a change of perspective on supersymmetric models.
If low energy supersymmetry is realized in Nature and if it is to be "natural", the simplest (universal) pattern  of soft supersymmetry breaking  terms looks less plausible. In  this paper we have explored the possibility of linking natural supersymmetry to flavour physics. We have revived the idea of obtaining natural  supersymmetry from a flavour theory based on a horizontal symmetry. The proposed model is based on spontaneously broken  $G\times U(1)_{local}$  family symmetry, where G is a discrete non-abelian subgroup of $SU(2)$, with both F-term and D-term supersymmetry breaking. \footnote{Another possibility is that  the $SU(2)$ group  is weakly gauged, with sufficiently suppressed  D-terms.}
The soft masses depend then on the two mass scales  and their flavour dependence is predicted as a consequence of the hierarchical pattern of fermion masses. A fit to fermion masses and mixings gives a small number of sets of the horizontal charges consistent with the data. The bounds on the sfermion sector  from the CP-violation in the kaon system and the other phenomenological constraints discussed in Section \ref{RG}  point then almost uniquely to the set with the same U(1) charges for all fermion fields except the left-handed top-bottom  quark   doublet and the right-handed top (with the interchange
left $\leftrightarrow$ right for   charged leptons, if the charges respect SU(5) symmetry), which remain zero.

The spectrum of sfermions appears as two sets of particles. The light one, with zero horizontal charges, obtain their mass from the F-term supersymmetry breaking. The dominant contribution to the mass of the heavy ones  comes from the D-term breaking.   Insisting, for naturalness, on the light masses to be in  the 1 TeV range, for the heavy masses one obtains a narrow range of a few to a few tens of TeV, with the lower bound determined by the CP-violation constraint and the upper bound given by the constraints on the left- versus right-handed sbottom mass splitting discussed in Section \ref{RG}. We emphasize  the latter point as a model independent constraint on the spectra of minimal  supersymmetry, in which only stops and the left-handed sbottom are  light. The required high scale hierarchy of $\frac{D}{ (F/M)^2}$  is  a factor of (3-5).

Besides linking natural supersymmetry to flavour physics, the model combines the advantages and  minimizes the disadvantages of the abelian $U(1)$ and non-abelian $U(2)$ symmetries as theories of flavour in supersymmetric models. For instance,
$U(2)$ models are in some tension with the fermions mass and mixing data. In particular, the relation $|V_{ub}/V_{cb}| \simeq \sqrt{{m_u}/{m_c}}$ predicted in this framework is off by a factor of two compared to the data.  We found an interesting indirect  correlation between correcting that result and  the  magnitude of the right-handed rotations in the (3,2) sector required to diagonalise the down quark Yukawa matrix. 
This, in turn has a strong impact on the bounds obtained from the FCNC and CP violation data on the masses of first two families of squarks and the right-handed sbottom mass.  Another point worth mentioning is that $U(2)$ models
predict generically large $\tan \beta$, which together with constraints from the RG running from a high (GUT) scale forces the right-handed sbottom to stay light,  
leading therefore to a non-minimal natural supersymmetry spectrum. 

The phenomenological success   of the model makes it worth exploring in detail the UV  mechanisms   of  the D-term supersymmetry breaking in the presence of symmetry groups beyond simple U(1), which have to be spontaneously broken with a proper alignment of the flavon vevs. In the present case these are either gauged $SU(2)$ with suppressed non-abelian D-terms or a discrete subgroup of global $SU(2)$. The latter may fit very well into the theoretical framework  of string theory. We discussed  general requirements for such a mechanism. 
\label{conclusions}

\vskip 1cm
%%%%%%%%%%%%%%%%%%%%%%%%%%%%%%%%%%%%%%%%%%%%%%%%%%%%%%%%%%%%%%%%%%%%%%%%%%
\section*{Acknowledgements}  
We would like to thank Marcin Badziak for help and collaboration in Section \ref{RG}. We are also grateful to L. Calibbi, M. Nardecchia, P. Paradisi and A. Romanino for useful discussions. E.D. and G.G. 
thank the Galileo Galilei Institute for Theoretical Physics for the hospitality and the INFN for partial support during the 
completion of this work. 
GG would like to thank the Funda\c{c}\~ao de
Amparo \`a Pesquisa do Estado de S\~ao Paulo (FAPESP) for financial support.
This  work was supported in part by the European ERC Advanced Grant 226371 MassTeV, the 
French ANR TAPDMS ANR-09-JCJC-0146 and the contract PITN-GA-2009-237920 UNILHC.
This work was also supported by National Science Centre in Poland under research 
grants DEC-2011/01/M/ST2/02466, DEC-2012/04/A/ST2/00099, 
DEC-2012/05/B/ST2/02597. Finally we acknowledge support from the ERC Advanced Grant project ``FLAVOUR" (267104) and the TUM Institute for Advanced Study.

%%%%%%%%%%%%%%%%%%%%%%%%%%%%%%%%%%%%%%%%%%%%%%%%%%%%%%%%%%%%%%%%%

\begin{appendix}
\section{Fermion Sector}
\label{fermionsector}

The Yukawa matrices as obtained from the superpotential in Eq.~(\ref{Wyuk}) are given by
\begin{align}
Y_u & = 
\begin{pmatrix}
0 & h_{12}^u \epsup & 0 \\
- h_{12}^u  \epsup  & h_{22}^u \epsu^2& h_{23}^u  \epsu \\
0 & h_{32}^u \epsu & h_{33}^u 
\end{pmatrix}, \\
Y_d & = 
\begin{pmatrix}
0 & h_{12}^d \epsup \epsd/\epsu & 0 \\
- h_{12}^d \epsup \epsd/\epsu & h_{22}^d \epsu \epsd & h_{23}^d  \epst \epsu \\
0 & h_{32}^d \epsd & h_{33}^d \epst 
\end{pmatrix}, %\\
%Y_e & = 
%\begin{pmatrix}
%0 & h_{12}^e \epsup \epsd/\epsu & 0 \\
%- h_{12}^e \epsup \epsd/\epsu & h_{22}^e \epsu \epsd & h_{23}^e  \epst \epsu \\
%0 & h_{32}^e \epsd & h_{33}^e \epst 
%\end{pmatrix}.
\end{align}
These matrices are diagonalized by unitary matrices $V^I_{L,R}$ according to
\begin{align}
\label{DiagYuk}
(V^I_L)^T \, Y_I \, V_R^I  = Y_I^{\rm diag}, \qquad I  = u,d,
\end{align}
such that the mass eigenstates are related to the original quarks (denoted by $'$)
\be
u_L'=V_L^u\, u_L\,,\qquad
d_L' =V_L^d\,d_L\,,\qquad
u_R'=V_R^{u*}u_R\,,\qquad
d_R'=V_R^{d*}d_R\,.
\label{quarkeigenstates}
\ee
The eigenvalues are approximately\footnote{All expressions receive multiplicative corrections of the form $(1 + \ord{\epsu^2, \epsup/\epsu}).$} given by
\begin{align}
Y_t & \approx |Y^u_{33}|, & Y_c & \approx \frac{|Y^u_{22}Y^u_{33}-Y^u_{23}Y^u_{32}|}{Y_t}, & Y_u & \approx  \frac{|(Y^u_{12})^2 Y^u_{33}|}{Y_tY_c} , \label{diagonalyukawasup}\\
Y_b & \approx  \sqrt{|Y^d_{33}|^2+|Y^d_{32}|^2} , & Y_s & \approx \frac{|Y^d_{22}Y^d_{33}-Y^d_{23}Y^d_{32}|}{Y_b}  ,  & Y_d & \approx  \frac{|(Y^d_{12})^2 Y^d_{33}|}{Y_bY_s}
 \, , \label{diagonalyukawasdown}%\\
%Y_\tau & \approx  \sqrt{|h^e_{33}\eps_3|^2+|h^e_{32}\eps_d|^2} , & y_\mu & \approx \frac{|h^e_{22}h^e_{33}-h^e_{23}h^e_{32}|}{y_\tau} \epsu \epsd\epst,  & y_e & \approx  \frac{|(h^e_{12})^2 h^e_{33}|}{y_\tau y_\mu}
%\frac{\epsu^{\prime 2} \epsd^2 \epst}{\epsu^2} \,.
%y_{\tau} & \approx  |h^e_{33}| \epst, & y_{\mu} & \approx |\frac{h^e_{22}h^e_{33}-h^e_{23}h^e_{32}}{h^e_{33}}| \epsu \epsd,  & y_e & \approx  |\frac{(h^e_{12})^2 h^e_{33}}{h^e_{22}h^e_{33}-h^e_{23}h^e_{32}}| \epsu^{\prime 2} \epsd/\epsu^3.  
\end{align}
The unitary rotations have the form 
\begin{align}
\label{rots}
V^{u,d}_L & = P'_L\, \hat{V}^{u,d}_L\, P_L \, , 
%& V^{e}_L & = \hat{V}^{e}_L P_L^e\, ,  
& V^{I}_R & = P'^I_R\, \hat{V}^{I}_R\, P^I_R\, , 
\end{align}
where $P^I_{L,R}$ and $P'^I_{L,R}$ are diagonal phase matrices to be fixed below, and $\hat{V}_{L,R}^I$ are approximately given by
\begin{align}
\hat{V}^u_L & = 
\begin{pmatrix}
1 & V^u_{12} & 0 \\
- V^{u*}_{12} & 1 & V^u_{23} \\
V^{u*}_{12} V^{u*}_{23}  & - V^{u*}_{23}  & 1
\end{pmatrix}, 
& \hat{V}^u_R & =  \begin{pmatrix}
1 & - V^{u}_{12} & 0 \\
V^{u*}_{12} & 1 & V^{u}_{32} \\
- V^{u*}_{12} V^{u*}_{32}  & - V^{u*}_{32}  & 1
\end{pmatrix},\\
\hat{V}^d_L & = 
\begin{pmatrix}
1 & V^d_{12} & V^d_{13} \\
- V^{d*}_{12} & 1 & V^d_{23} \\
V^{d*}_{23} V^{d*}_{12} - V^{d*}_{13} & - V^{d*}_{23}  & 1\end{pmatrix}, 
 & \hat{V}^d_R & =  \begin{pmatrix}
1 & - V^{d}_{12}/c_d & 0 \\
V^{d*}_{12} & c_d & V^{d}_{32} \\
- V^{d*}_{12} V^{d*}_{32}/c_d  & - V^{d*}_{32}  & c_d
\end{pmatrix},%\\
% \hat{V}^e_L & = 
%\begin{pmatrix}
%1 & V^e_{12} & V^e_{13} \\
%- V^{e*}_{12} & 1 & V^e_{23} \\
% V^{e*}_{23} V^{e*}_{12} - V^{e*}_{13}  & - V^{e*}_{23}  & 1\end{pmatrix}, 
% & \hat{V}^e_R & =  \begin{pmatrix}
%1 & - V^{e}_{12}/c_e & 0 \\
%V^{e*}_{12} & c_e & V^{e}_{32} \\
%- V^{e*}_{12} V^{e*}_{32}/c_e  & - V^{e*}_{32}  & c_e
%\end{pmatrix} \ , 
\end{align}
with
\begin{align}
V^u_{12}& \approx \left( \frac{Y^u_{12}Y^u_{33}}{Y^u_{22}Y^u_{33}-Y^u_{23}Y^u_{32}} \right)^*  \equiv |V^u_{12}|  e^{i \alpha^u_{12}} =  e^{i \alpha^u_{12}} \sqrt{\frac{m_u}{m_c}}  \ , \\
%\end{align}
%\begin{align}
V^u_{23}&\approx\frac{Y^{u*}_{23} Y^{u}_{33}}{Y_t^2}  \equiv e^{i \alpha^u_{23}} |V^u_{23}|   \, ,  \\
 V^{u}_{32}&\approx\frac{Y^{u*}_{32} Y^{u}_{33}}{Y_t^2}  \equiv e^{i \alpha^u_{32}} |V^u_{32}|   \, , \\
% V^u_{13}&\approx \max \mathcal O(\epsu \epsu', \epsilon_{\phi} \epsilon_{\chi}^{X_{10}-X_{\phi}})\, , 
%\frac{h^u_{12}h^{u*}_{32}}{y_t^2}\epsu' \epsu
%\end{align}
%\begin{align}
V^d_{12}&\approx \left( \frac{Y^d_{12}Y^d_{33}}{Y^d_{22}Y^d_{33}-Y^d_{23}Y^d_{32}} \right)^*  \equiv e^{i \alpha^d_{12}}|V^d_{12}|   = e^{i \alpha^d_{12}} \sqrt{\frac{m_d}{m_s}} \sqrt{c_d}   \,, \\ 
V^d_{13}&\approx\frac{Y^{d*}_{12}Y^{d}_{32}}{Y_b^2} \equiv e^{i \alpha^d_{13}} |V^d_{13}|   = e^{i \alpha^d_{13}}  \sqrt{\frac{m_d m_s}{ m_b^2}} \frac{s_d}{\sqrt{c_d}}  \, , \\
%\end{align}
%\begin{align}
V^d_{23}&\approx\frac{Y^{d*}_{23}Y^{d}_{33}+Y^{d*}_{22}Y^{d}_{32}}{Y_b^2} 
\equiv e^{i \alpha^d_{23}}|V^d_{23}|  \, ,\\ 
 V^d_{32}&\approx\frac{Y^{d*}_{32}}{Y^{d*}_{33} } \frac{|Y^d_{33}| }{Y_b} \equiv e^{i \alpha^d_{32}}|V^d_{32}|   = e^{i \alpha^d_{32}} s_d  \,
\label{largeRHangle}
\end{align}
%\begin{align}
%V^e_{12}&\approx \left( \frac{h^e_{12}h^e_{33}}{h^e_{22}h^e_{33}-h^e_{23}h^e_{32}} \right)^* \frac{\epsu'}{\epsu^2} \equiv |V^e_{12}|  e^{i \alpha^e_{12}} = e^{i \alpha^e_{12}} \sqrt{m_e/m_\mu} \sqrt{c_e}   \,, \\ 
%V^e_{13}&\approx\frac{h^{e*}_{12}h^{e}_{32}\epsd^2}{y_\tau^2} \frac{\epsu'}{\epsu}\equiv |V^e_{13}|  e^{i \alpha^e_{13}} = e^{i \alpha^e_{13}}  \sqrt{m_e m_\mu/ m_\tau^2} \frac{s_e}{\sqrt{c_e}}  \, , 
%\end{align}
%\begin{align}
%V^e_{23}&\approx\frac{h^{e*}_{23}h^{e}_{33}\epst^2+h^{e*}_{22}h^{e}_{32}\epsd^2}{y_\tau^2}\epsu \equiv |V^e_{23}|  e^{i \alpha^e_{23}}\, , 
%& V^e_{32}&\approx\frac{h^{e*}_{32}}{h^{e*}_{33} } \frac{|h^e_{33}| \epsd}{y_\tau} \equiv |V^e_{32}|  e^{i \alpha^e_{32}} = e^{i \alpha^e_{32}} s_e.
%\end{align}
Here we neglected contribution to the 1-3 entries of the rotations that are of order 
\begin{align}
(\hat{V}^u_{R,L})_{13} & = \max  \ord{\epsu \epsu^\prime,\epsilon_{\phi} \epsilon_{\chi}^{X_{10}-X_{\phi}}}\, , 
&  (\hat{V}^{d}_R)_{13} & = \max  \ord{\epsu^\prime,\epsilon_{\phi} \epsilon_{\chi}^{X_{53}-X_{\phi}}}\, , 
\end{align}
and we defined $c_{d} \equiv \cos \theta_{d} $, $s_{d} \equiv \sin \theta_{d} $, where the angle $\theta_{d}$ is defined by
\begin{align}
\tan \theta_{d} & \equiv  \frac{|Y^d_{32}|}{|Y^d_{33}|}  \, , 
%&\tan \theta_{e} & \equiv  \frac{|h^e_{32}|}{|h^e_{33}|} \frac{\epsd}{\epst} \, . 
\end{align} 
This angle parametrizes the all-order corrections in $\epsd/\epst$, which typically is not a very small expansion parameter. The twelve angles present in a general set of quark rotations are thus reduced to only four, given by the free parameters $|V_{23}^{u,d}|$ and $|V_{32}^{u,d}|$. This is due to the three textures zeroes and the relation $Y_{12}=-Y_{21}$ in the Yukawas, which indeed remove eight real degrees of freedom. Below we will see that the phase structure is also greatly simplified, as there are only four phases that can be physically relevant.
Matching this to the degees of freedom in the CKM matrix, we see that we have one angle and three phases as free parameters.

The left-handed phases matrices can be chosen to bring the CKM matrix to the standard PDG form with one physical phase, while the right-handed ones render the diagonal Yukawas real and positive. This gives
\begin{align}
P_L & = \begin{pmatrix} 
 e^{i (\tilde{\alpha}_{12}+\alpha^d_{12}  ) } & 0 & 0 \\
 0 & 1  & 0 \\
 0 & 0 & e^{- i (\tilde{\alpha}_{23}+\alpha^d_{23}  )}
 \end{pmatrix}\, , 
 \end{align}
 \begin{align}
 P_R^u & =  e^{- i \arg m_c}  \begin{pmatrix} 
 e^{- i (\tilde{\alpha}_{12}+\alpha_{12}  - \alpha^u_{12})} & 0 & 0 \\
 0 & 1  & 0 \\
 0 & 0 & e^{ i (\tilde{\alpha}_{23} +\alpha^d_{23} +  \arg m_c - \arg m_t)}
 \end{pmatrix}\, , \\ 
  P_R^d & = e^{- i \arg m_s} \begin{pmatrix} 
 e^{- i ( \tilde{\alpha}_{12} - \alpha^d_{12})} & 0 & 0 \\
 0 & 1  & 0 \\
 0 & 0 & e^{ i ( \tilde{\alpha}_{23}+\alpha^d_{23}+\arg m_s-\arg m_b  )}
 \end{pmatrix}\, , 
\end{align} 
with 
\begin{align}
\alpha_{12} & \equiv \alpha^d_{12} - \alpha^u_{12} \, , & 
\alpha_{23} & \equiv \alpha^d_{23} - \alpha^u_{23} \, , 
\end{align}
\begin{align}
\tilde{\alpha}_{12} & = \arg \left( 1 - \frac{|V^u_{12}|}{|V^d_{12}|}  e^{-i \alpha_{12}} \right) \, , &
\tilde{\alpha}_{23} & = \arg \left( 1 - \frac{|V^u_{23}|}{|V^d_{23}|} e^{-i \alpha_{23}} \right) \, , 
\end{align}
and we used the fact that under the rotations $\hat V_{L,R}^{u,d}$ the eigenvalues have phases given by
\begin{align}
\arg m_t & = \arg h^u_{33} \, , & \arg m_c & = \arg h^u_{12} + \alpha^u_{12} \, , &\arg m_u & = - 2 \, \alpha^u_{12} + \arg m_c\, \\
\arg m_b & = \arg h^d_{33} \, , & \arg m_s & = \arg h^d_{12} + \alpha^d_{12} \, , & \arg m_d & = - 2 \, \alpha^d_{12} + \arg m_s.
\end{align}
%and we used the relations
%
%\begin{align} 
%\arg m_u & = - 2 \, \alpha^u_{12} + \arg m_c\, , & \arg m_d & = - 2 \, \alpha^d_{12} + \arg m_s.
%\end{align}
%
For later use, we will also make the definitions
\begin{align}
\alpha_u&\equiv\arg m_t-\arg m_c-\alpha^u_{32}-\alpha_{23}^u\\
\alpha_d&\equiv\arg m_b-\arg m_s-\alpha^d_{32}-\alpha_{23}^d\nonumber\\
&=\alpha^d_{13}-\alpha^d_{23}-\alpha^d_{12}
\end{align}
Finally we can mutliply our rotations $\hat V_{L,R}^{u,d}$ by phases $P'^I_{R}$, $P'_L$ from the left without any physical effect (as we can absorb these phases by a redefinition of the original fields $q',\ u',\ $ and $d'$). Using this freedom, we can bring the quark rotations to their final form:
%
%\be
%V_L^d=
%\begin{pmatrix}
%1	&|V_{12}^d|\,e^{-i\tilde \alpha_{12}}&|V^d_{13}|\,e^{-i(\tilde\alpha_{12}+\tilde \alpha_{23}-\alpha_d)}\\
%-|V_{12}^d|\,e^{i\tilde \alpha_{12}}	& 1	&|V_{23}^d|\,e^{-i\tilde \alpha_{23}}\\
%(|V_{12}^dV_{23}^d|-|V^d_{13}|\,e^{-i\alpha_d})\,e^{i(\tilde \alpha_{12}+\tilde\alpha_{23})}&-|V_{23}^d|\,e^{i%\tilde \alpha_{23}}&1
%\end{pmatrix}
%\ee
\be
V_L^d=
\begin{pmatrix}
1	&|V_{12}^d|&|V^d_{13}|\,e^{i\alpha_d}\\
-|V_{12}^d|	& 1	&|V_{23}^d|\\
|V_{12}^dV_{23}^d|-|V^d_{13}|\,e^{-i\alpha_d}&-|V_{23}^d|&1
\end{pmatrix}
\begin{pmatrix}
e^{i\tilde\alpha_{12}}&&\\
&1&\\
&&e^{-i\tilde\alpha_{23}}
\end{pmatrix}\,,
\ee
%\be
%V_L^u=
%\begin{pmatrix}
%1	&|V_{12}^u|\,e^{-i\hat \alpha_{12}}&0\\
%-|V_{12}^u|\,e^{i\hat \alpha_{12}}	& 1	&|V_{23}^u|\,e^{-i\hat \alpha_{23}}\\
%|V_{12}^uV_{23}^u|\,e^{i(\hat \alpha_{12}+\hat\alpha_{23})}&-|V_{23}^u|\,e^{i\hat \alpha_{23}}&1
%\end{pmatrix}
%\ee
%
%
\be
V_L^u=
\begin{pmatrix}
e^{-i\alpha_{12}}\\&1\\&&e^{i\alpha_{23}}
\end{pmatrix}
\begin{pmatrix}
1	&|V_{12}^u|&0\\
-|V_{12}^u|	& 1	&|V_{23}^u|\\
|V_{12}^uV_{23}^u|&-|V_{23}^u|&1
\end{pmatrix}
\begin{pmatrix}
e^{i(\tilde\alpha_{12}+\alpha_{12})}&&\\
&1&\\
&&e^{-i(\tilde\alpha_{23}+\alpha_{23})}
\end{pmatrix}\,,
\ee
%\be
%V_R^d=
%\begin{pmatrix}
%1	&-|V_{12}^d|/c_d\,e^{i\tilde \alpha_{12}}&	0\\
%|V_{12}^d|\,e^{-i\tilde \alpha_{12}}	& c_d	&|V_{32}^d|\,e^{i(\tilde \alpha_{23}-\alpha_d)}\\
%-|V_{12}^dV_{32}^d|/c_d\,e^{-i(\tilde \alpha_{12}+\tilde\alpha_{23}-\alpha_d)}&-|V_{32}^d|\,e^{-i(\tilde \alpha_{23}-\alpha_d)}& c_d
%\end{pmatrix}
%\ee
\be
V_R^d=
\begin{pmatrix}
1	&-|V_{12}^d|/c_d&	0\\
|V_{12}^d|	& c_d	&|V_{32}^d|\\
-|V_{12}^dV_{32}^d|/c_d&-|V_{32}^d|& c_d
\end{pmatrix}
\begin{pmatrix}
e^{-i\tilde\alpha_{12}}&&\\
&1&\\
&&e^{i(\tilde\alpha_{23}-\alpha_{d})}
\end{pmatrix}\,,
\ee
%
%
%\be
%V_R^u=
%\begin{pmatrix}
%1	& -|V_{12}^u|\,e^{i\hat \alpha_{12}}&0\\
%%|V_{12}^u|\,e^{-i(\tilde \alpha_{12}+\alpha_{12})}	& 1	&|V_{32}^u|\,e^{i(\hat \alpha_{23}-\alpha_u)}\\
%-|V_{12}^uV_{32}^u|\,e^{-i(\hat \alpha_{12}+\hat\alpha_{23}-\alpha_u)}&-|V_{32}^u|\,e^{-i(\hat \alpha_{23}-\alpha_u)}&1
%\end{pmatrix}
%\ee
\be
V_R^u=
\begin{pmatrix}
1	& -|V_{12}^u|&0\\
|V_{12}^u|	& 1	&|V_{32}^u|\\
-|V_{12}^uV_{32}^u|&-|V_{32}^u|&1
\end{pmatrix}
\begin{pmatrix}
e^{-i(\tilde\alpha_{12}+\alpha_{12})}&&\\
&1&\\
&&e^{i(\tilde\alpha_{23}+\alpha_{23}-\alpha_{u})}
\end{pmatrix}\,.
\ee
Any physical observable can thus only depend on the four phases $\alpha_{12}$ (also through $ \tilde\alpha_{12}$), 
$\alpha_{23}$ (also through $\tilde\alpha_{23}$), $\alpha_u$ and $\alpha_d$.
The CKM matrix $V_{CKM} = (V_L^u)^\dagger V_L^d  $ is then given by 
\be
V_{CKM}  = \begin{pmatrix}
1 & |V_{12}| &  |V_{13}| e^{- i \delta } \\
- |V_{12}| & 1 & |V_{23}| \\
 |V_{12} V_{23} | - |V_{13}| e^{i \delta} & - |V_{23}| & 1
\end{pmatrix}
\ee
with 
\begin{align}
V_{12} & = V^d_{12} - V^u_{12}\, ,  & 
V_{23} & = V^d_{23} - V^u_{23}\, ,  \\
V_{13} & = V^d_{13} - V^u_{12} V_{23}\, ,  & 
| V_{td} | & = | V_{12} V_{23} - V_{13} |  \, ,  
\end{align}
and the CKM phase is given by
\begin{align}
\delta = \alpha_{12} + \tilde{\alpha}_{12} - \arg \left(  \frac{| V^d_{13} |}{|V^u_{12} V_{23}| } e^{i ( \alpha_d +\alpha_{12} - \tilde{\alpha}_{23})} -1 \right)\, .  
\end{align}
The above form of the CKM matrix gives rise to the relations
\begin{align}
|V_{us}|  & \approx \sqrt{m_d/m_s} \sqrt{c_d}\,,
\end{align}
\begin{align}
  |V_{ub}/V_{cb}|  & \approx  |\sqrt{m_u/m_c}+e^{i\beta}\Delta \,t_d \sqrt{c_d}| \ , &
     |V_{td}/V_{ts}|  & \approx | \sqrt{m_d/m_s}+e^{i\beta'}\Delta \ t_d |\sqrt{c_d} \ ,       
\end{align}
where $t_d \equiv \tan \theta_d$ and 
\begin{align}
\Delta & =\frac{ \sqrt{m_s m_d}}{|V_{cb}| m_b} \approx0.09 \ ,
& \beta & = \alpha_d +\alpha_{12} - \tilde{\alpha}_{23} + \pi \ , & \beta^\prime & = \beta - \alpha_{12} \ .
\label{Deltabeta}
\end{align}

The eigenvalues and rotations for the charged lepton sector can be
obtained by replacing everywhere in the above the index $d$ by $e$, with the
exception of the phases, which in the leptonic sector depend on the model of
neutrino masses.  
%%%%%%%%%%%%%%%%%%%%%%%%%%%%%%%%%%%%%%%%%%%%%%%%%%%%%%%%%%%%%%

\section{SUSY Contributions to Flavor Observables}
\label{wilsoncoefficients}

The new SUSY contributions to FCNC processes due to squark-gluino loops have been calculated in the literature in terms of the flavor-changing unitary matrices $Z_{U,D}$ appearing at the gluino vertex in  the notation of Ref.~\cite{Rosiek}
\be
\mathcal L = 
	\tilde u_L^* \,\bar\lambda\, (Z_U^{L})^\dagger u_L
	+\tilde u_R^*\,\bar\lambda\,(Z_U^R)^\dagger u_R
	+\tilde d_L^*\,\bar\lambda\,(Z_{D}^{L})^T d_L
	+\tilde d_R^*\,\bar\lambda\,(Z_D^R)^T d_R+h.c.
\ee
where we have used that the $LR$ flavour violation is negligible because of the suppression of the $A$ terms.
The flavour violation in the Kaon sector can then be encoded in the Wilson coefficients of the operators
\begin{align}
Q_1&=\left(\bar d_L\gamma_\mu s_L \right)^2\,,
&Q_2&=\left(\bar d_R  s_L \right)^2\,,
&Q_3&=\bar d_R^\beta s_L^\alpha \ \bar d_R^\alpha s_L^\beta \,,\\
\tilde Q_1&=\left(\bar d_R\gamma_\mu s_R \right)^2\,,
&\tilde Q_2&=\left(\bar d_L  s_R \right)^2\,,
&\tilde Q_3&=\bar d_L^\beta s_R^\alpha \ \bar d_L^\alpha s_R^\beta \,,
\end{align}
\be
Q_4=(\bar d_R s_L)\,(\bar d_L s_R)\,,\qquad  
Q_5=\bar d_R^\beta s_L^\alpha\,\bar d_L^\alpha s_R^\beta\,,
\ee
 For example the SUSY contribution to the Wilson coefficient of the effective operator $Q_1$, $\tilde Q_1$, $Q_4$ and $Q_5$ is given by~\cite{hagelin} 
\begin{align}
\label{C1general}
\Delta C_1 &= \frac{\alpha_s^2}{m_{\tilde{g}}^2}  (Z_D^L)_{1i}^* (Z_D^L)_{2i}   (Z_D^L)_{1j}^* (Z_D^L)_{2j} \left( \frac{11}{36} \tilde{f}_4 (x_i^L,x_j^L) + \frac{1}{9} f_4(x_i^L,x_j^L) \right)  \, ,\\
\Delta \tilde C_1 &= \frac{\alpha_s^2}{m_{\tilde{g}}^2}  (Z_D^R)_{1i}^* (Z_D^R)_{2i}   (Z_D^R)_{1j}^* (Z_D^R)_{2j} \left( \frac{11}{36} \tilde{f}_4 (x_i^R,x_j^R) + \frac{1}{9} f_4(x_i^R,x_j^R) \right)  \, ,\\
\Delta C_4 & = \frac{\alpha_s^2}{m_{\tilde{g}}^2} \,  
 (Z_D^L)_{1i}^* (Z_D^L)_{2i}   (Z_D^R)_{1j}^* (Z_D^R)_{2j}
 \left(- \frac{1}{3}  \tilde{f}_4 (x_i^L,x_j^R)  + \frac{7}{3} f_4 (x_i^L,x_j^R)
  \right)\,,\label{C4exact}\\
\Delta C_5 & = \frac{\alpha_s^2}{m_{\tilde{g}}^2} \,  
 (Z_D^L)_{1i}^* (Z_D^L)_{2i}   (Z_D^R)_{1j}^* (Z_D^R)_{2j}
 \left( \frac{5}{9}  \tilde{f}_4 (x_i^L,x_j^R)  + \frac{1}{9} f_4 (x_i^L,x_j^R)
  \right)  
\end{align} 
where $f_4, \tilde{f}_4$ are some loop functions given below and $x_i^L = m^2_{d^i_L}/m_{\tilde{g}}^2$, where $m_{\tilde{g}}$ is the gluino mass and $m_{d_L^i}$ is the mass of the d-squark.% Here we have assumed that the LR contribution to sfermion masses is negligible and used the notation for $Z_D$ of the previous section.

In the limit where the first two sfermion generations are degenerate, we can carry out the summation over $i,j$ and use unitarity of $Z^L_D$ to obtain a compact form for the Wilson coefficient in which the flavor suppression and the loop functions are factorized, e.g.
\begin{align}
\Delta C_1 & = \frac{\alpha_s^2}{m_{\tilde{g}}^2}  \left[ (Z_D^L)_{13}^* (Z_D^L)_{23}  \right]^2 \left[ \frac{11}{36} \left( \tilde{f}_4 (x_1^L,x_1^L) - 2 \tilde{f}_4 (x_1^L,x_3^L) + \tilde{f}_4 (x_3^L,x_3^L)\right) + \frac{1}{9} \left( \tilde{f}_4 \to f_4 \right) \right]  \, .
\end{align} 
The suppression of the $i \to j$ flavor transition is therefore entirely determined by the factor  $ (Z_D^L)_{i3}^* (Z_D^L)_{j3} $. 
In principle the unitary matrices $Z_I$ are a product of quark and squark rotations. In the case where all flavour violation comes from the quark rotations we can simply read them off Eq.~(\ref{quarkeigenstates}),
\begin{align}
\label{gluinovertexFV}
Z_U^L & = (V^u_L)^\dagger \, , & Z_U^R & = (V^u_R)^T \, , \\
Z_D^L & = (V^d_L)^T \, , & Z_D^R & = (V^d_R)^\dagger \, . 
\end{align}
 which in turn only depend on the unitary superfield rotations $V^d_L$. We therefore define the quantities\footnote{The analogue expressions of $\Delta C_1$ in the up sector are obtained with $Z_D \to Z_U^*$.} 
\begin{align}
\hat{\delta}^{d,LL}_{ij} & \equiv (Z_D^L)_{i3}^* (Z_D^L)_{j3} =  (V^d_L)_{3i}^*  (V^d_L)_{3j} \, ,  
& \hat{\delta}^{d,RR}_{ij} & \equiv (Z_D^R)_{i3}^* (Z_D^R)_{j3} = (V^d_R)_{3i}  (V^d_R)^*_{3j}\, , \\
\hat{\delta}^{u,LL}_{ij} & \equiv (Z_U^L)_{i3} (Z_U^L)_{j3}^* =  (V^u_L)_{3i}^*  (V^u_L)_{3j} \, ,  
& \hat{\delta}^{u,RR}_{ij} & \equiv (Z_U^R)_{i3} (Z_U^R)_{j3}^* = (V^u_R)_{3i}  (V^u_R)_{3j}^* \, .
\end{align}
Using the explicit expressions for $V^{u,d}_{L,R}$ in Appendix A, we obtain 
\begin{align}
\hat{\delta}^{d,LL}_{23} & = - |V^d_{23}| \,  e^{- i \tilde{\alpha}_{23}}, \\
 \hat{\delta}^{d,LL}_{13} & = \left( |V^d_{23} | \sqrtd \sqrt{c_d} - \sqrt{\frac{m_d m_s}{m_b^2}} \frac{s_d}{\sqrt{c_d}} e^{ i \alpha_d} \right) \, e^{- i (\tilde{\alpha}_{23} + \tilde{\alpha}_{12} )}, \\
\hat{\delta}^{d,LL}_{12} & = \left( - |V^d_{23}|^2 \sqrtd \sqrt{c_d} + \sqrt{\frac{m_d m_s}{m_b^2}} \frac{s_d}{\sqrt{c_d}} |V^d_{23}| e^{i  \alpha_d} \right) e^{-i \tilde{\alpha}_{12}}, \\
\hat{\delta}^{d,RR}_{23} & = - s_d \, c_d \, e^{-i \left(  \tilde{\alpha}_{23} -\alpha_d \right)}, \\
 \hat{\delta}^{d,RR}_{13} & = -  \sqrtd \sqrt{c_d} \, s_d e^{-i \left( \tilde{\alpha}_{12}   + \tilde{\alpha}_{23} -\alpha_d \right)}, \\
\hat{\delta}^{d,RR}_{12} & = \sqrtd \frac{s_d^2}{\sqrt{c_d}} e^{-i \tilde{\alpha}_{12}}.
\end{align}
%with the definition
%
%%\be
%\alpha^d\equiv \alpha^d_{13} - \alpha^d_{23} - \alpha^d_{12}
%\ee
%
%As one can see also directly from the initial Yukawa couplings, there are four fundamental phases in the quark sector ($\tilde{\alpha}_{12}, \tilde{\alpha}_{23},  \tilde{\alpha}_{U}, \tilde{\alpha}_{D}$) and one phase 
%in the lepton sector ($\tilde{\alpha}_E$). Indeed one can show that these phases are invariant under phase redefinitions of the initial parameters $h^{U,D,E}_{ij}$. 
Notice that the phase $\tilde{\alpha}_{12}$ that appears in the LL and RR 1-2 sector is small:
\be
\label{phase12}
\tilde{\alpha}_{12} =  \sqrt{\frac{m_u m_s}{m_d m_c}} \sin \alpha_{12} + \ord{\frac{m_u m_s}{m_d m_c}} \lesssim 0.2.
\ee
In terms of these quantities the relevant Wilson coefficients are given by
\begin{align}
\label{wilsonc1}
\Delta C_1 & = \frac{\alpha_s^2}{m_{\tilde{g}}^2}   \left( \hat{\delta}^{LL}_{12}  \right)^2 \left[ \frac{11}{36} \left( \tilde{f}_4 (x_1^L,x_1^L) - 2 \tilde{f}_4 (x_1^L,x_3^L) + \tilde{f}_4 (x_3^L,x_3^L)\right) + \frac{1}{9} \left( \tilde{f}_4 \to f_4 \right) \right]  \, , \\
\label{wilsonc1t}
\Delta \tilde{C}_1 & = \frac{\alpha_s^2}{m_{\tilde{g}}^2}   \left( \hat{\delta}^{RR}_{12}  \right)^2 \left[ \frac{11}{36} \left( \tilde{f}_4 (x_1^R,x_1^R) - 2 \tilde{f}_4 (x_1^R,x_3^R) + \tilde{f}_4 (x_3^R,x_3^R)\right) + \frac{1}{9} \left( \tilde{f}_4 \to f_4 \right) \right]  \, , \\
\label{wilsonc4}
\Delta C_4 & = \frac{\alpha_s^2}{m_{\tilde{g}}^2} \,  \hat{\delta}^{LL}_{12}   \hat{\delta}^{RR}_{12}  \left[ - \frac{1}{3} \left( \tilde{f}_4 (x_1^L,x_1^R) -  \tilde{f}_4 (x_1^L,x_3^R) -   \tilde{f}_4 (x_3^L,x_1^R) + \tilde{f}_4 (x_3^L,x_3^R)\right) + \frac{7}{3} \left( \tilde{f}_4 \to f_4 \right) \right]  \, , \\
\label{wilsonc5}
\Delta C_5 & = \frac{\alpha_s^2}{m_{\tilde{g}}^2} \,  \hat{\delta}^{LL}_{12}   \hat{\delta}^{RR}_{12}  \left[ \frac{5}{9} \left( \tilde{f}_4 (x_1^L,x_1^R) -  \tilde{f}_4 (x_1^L,x_3^R) -  \tilde{f}_4 (x_3^L,x_1^R) + \tilde{f}_4 (x_3^L,x_3^R)\right) + \frac{1}{9} \left( \tilde{f}_4 \to f_4 \right) \right]  \, ,
\end{align}
with the loop functions
\begin{align}
\label{loopfunctions1}
f_4(x,x) & = \frac{2-2x + (1+x) \log x}{(x-1)^3} \, ,\\
f_4(x,y) & = \frac{x (y-1)^2 \log x - y (x-1)^2 \log y - (x-1) (y-1) (y-x)}{(x-1)^2 (y-1)^2 (y-x)} \, ,\\
\tilde{f}_4(x,x) & = \frac{1-x^2 + 2 x \log x}{(x-1)^3} \, ,\\
\tilde{f}_4(x,y) & = \frac{x^2 (y-1)^2 \log x - y^2 (x-1)^2 \log y - (x-1) (y-1) (y-x)}{(x-1)^2 (y-1)^2 (y-x)} \, .
\label{loopfunctions2}
\end{align}
Finally we give the loop function used for the calculation of BR $(\mu \to e \gamma)$
\be
f(\mu, M_2, \tan \beta, \tilde{m}^2_{\nu_k}) = f^c_R (x_{2 k}) + \frac{\mu (\mu + M_2 \tan \beta)}{M_2^2 - \mu^2} f^c_{LR}(x_{\mu k}) -  \frac{M_2 (M_2 + \mu \tan \beta)}{M_2^2 - \mu^2} f^c_{LR}(x_{2 k}) \, ,
\ee
where
\begin{align}
f^c_L(x) & = \frac{2 +3 x - 6 x^2 +  x^3 + 6 x \log x }{6 (1-x)^4} \, , & f^c_{LR}(x) & = \frac{-3 + 4x   -x^2 -  2  \log x }{(1-x)^3} \, 
\end{align}
and  $x_{2 k} = M_2^2/\tilde{m}^2_{\nu_k}$, $x_{\mu k} = 
\mu^2/\tilde{m}^2_{\nu_k}$.
%%%%%%%%%%%%%%%%%%%%%%%%%%%%%%%%%%%%%%%%%%%%%%%%%%%%%%%%%%%%%%%%%%%%%%%%%%%%%
\end{appendix}

%%%%%%%%%%%%%%%%%%%%%%%%%%%%%%%%%%%%%%%%%%%%%%%%%%%%%%%%%%%%%%%%%%%


\begin{thebibliography}{99}


\bibitem{CKN} 
  A.~G.~Cohen, D.~B.~Kaplan and A.~E.~Nelson,
  %``The More minimal supersymmetric standard model,''
  Phys.\ Lett.\ B {\bf 388}, 588 (1996)
  [hep-ph/9607394].
  %%CITATION = HEP-PH/9607394;%%
  %478 citations counted in INSPIRE as of 25 Jun 2013
 

\bibitem{SUSYATLAS} 
  G.~Aad {\it et al.}  [ATLAS Collaboration],
  %``Search for squarks and gluinos with the ATLAS detector in final states with jets and missing transverse momentum using 4.7 fb^{-1}$ of $\sqrt{s}=7$ TeV proton-proton collision data,''
  Phys.\ Rev.\ D {\bf 87}, 012008 (2013)
  [arXiv:1208.0949 [hep-ex]].
  %%CITATION = ARXIV:1208.0949;%%
  %101 citations counted in INSPIRE as of 25 Jun 2013
  
 \bibitem{SUSYCMS} 
  S.~Chatrchyan {\it et al.}  [CMS Collaboration],
  %``Search for supersymmetry in hadronic final states using MT2 in $pp$ collisions at $\sqrt{s} = 7$ TeV,''
  JHEP {\bf 1210}, 018 (2012)
  [arXiv:1207.1798 [hep-ex]].
  %%CITATION = ARXIV:1207.1798;%%
  %59 citations counted in INSPIRE as of 25 Jun 2013 

\bibitem{Perez1}
  R.~Mahbubani, M.~Papucci, G.~Perez, J.~T.~Ruderman and A.~Weiler,
  %``Light non-degenerate squarks at the LHC,''
  Phys.\ Rev.\ Lett.\  {\bf 110} (2013) 151804
  [arXiv:1212.3328 [hep-ph]];
  %%CITATION = ARXIV:1212.3328;%%
  %11 citations counted in INSPIRE as of 05 Sep 2013
  M.~Blanke, G.~F.~Giudice, P.~Paradisi, G.~Perez and J.~Zupan,
  %``Flavoured Naturalness,''
  JHEP {\bf 1306} (2013) 022
  [arXiv:1302.7232 [hep-ph]];
  %%CITATION = ARXIV:1302.7232;%%
  %10 citations counted in INSPIRE as of 05 Sep 2013
  I.~Galon, G.~Perez and Y.~Shadmi,
  %``Non-Degenerate Squarks from Flavored Gauge Mediation,''
  arXiv:1306.6631 [hep-ph].
  %%CITATION = ARXIV:1306.6631;%%
  %1 citations counted in INSPIRE as of 05 Sep 2013

\bibitem{Csaki}
T.~Gherghetta, B.~von Harling and N.~Setzer,
  %``A natural little hierarchy for RS from accidental SUSY,''
  JHEP {\bf 1107} (2011) 011
  [arXiv:1104.3171 [hep-ph]];
  %%CITATION = ARXIV:1104.3171;%%
C.~ Csaki, L.~ Randall and J.~Terning,
Phys.Rev. D86 (2012) 075009
arXiv:1201.1293 [hep-ph];
G. Larsen, Y. Nomura, H.L.L. Roberts, 
JHEP 1206 (2012) 032
arXiv:1202.6339 [hep-ph]; 
N.~Craig, M.~McCullough and J.~Thaler,
  %``The New Flavor of Higgsed Gauge Mediation,''
  JHEP {\bf 1203} (2012) 049
  [arXiv:1201.2179 [hep-ph]] and
  %%CITATION = ARXIV:1201.2179;%%
  %``Flavor Mediation Delivers Natural SUSY,''
  JHEP {\bf 1206} (2012) 046
  [arXiv:1203.1622 [hep-ph]];
  %%CITATION = ARXIV:1203.1622;%%
 R.~Auzzi, A.~Giveon, S.~B.~Gudnason and T.~Shacham,
  %``A Light Stop with Flavor in Natural SUSY,''
  JHEP {\bf 1301} (2013) 169
  [arXiv:1208.6263 [hep-ph]].
  %%CITATION = ARXIV:1208.6263;%%

\bibitem{ruderman}
  M.~Papucci, J.~T.~Ruderman and A.~Weiler,
  %``Natural SUSY Endures,''
  JHEP {\bf 1209} (2012) 035
  [arXiv:1110.6926 [hep-ph]];
  %%CITATION = ARXIV:1110.6926;%%
C.~Brust, A.~Katz, S.~Lawrence and R.~Sundrum,
  %``SUSY, the Third Generation and the LHC,''
  JHEP {\bf 1203} (2012) 103
  [arXiv:1110.6670 [hep-ph]];
  %%CITATION = ARXIV:1110.6670;%%
   H.~Baer, V.~Barger, P.~Huang and X.~Tata,
  %``Natural Supersymmetry: LHC, dark matter and ILC searches,''
  JHEP {\bf 1205} (2012) 109
  [arXiv:1203.5539 [hep-ph]].
  %%CITATION = ARXIV:1203.5539;%%
 
 \bibitem{AnomalousU1}
  P.~Binetruy and E.~Dudas,
  %``Gaugino condensation and the anomalous U(1),''
  Phys.\ Lett.\ B {\bf 389} (1996) 503
  [hep-th/9607172];
  %%CITATION = HEP-TH/9607172;%%
 G.~R.~Dvali and A.~Pomarol,
  %``Anomalous U(1) as a mediator of supersymmetry breaking,''
  Phys.\ Rev.\ Lett.\  {\bf 77} (1996) 3728
  [hep-ph/9607383].
  %%CITATION = HEP-PH/9607383;%%

\bibitem{Arkani}
 N.~Arkani-Hamed and H.~Murayama,
 Phys.Rev. D{\bf 56} (1997) 6733
 [arXiv:hep-ph/9703259];
 H.~Baer, V.~Barger, P.~Huang and X.~Tata,
  %``Natural Supersymmetry: LHC, dark matter and ILC searches,''
  JHEP {\bf 1205} (2012) 109
  [arXiv:1203.5539 [hep-ph]];
  %%CITATION = ARXIV:1203.5539;%%
F.~Brummer, S.~Kraml and S.~Kulkarni,
  %``Anatomy of maximal stop mixing in the MSSM,''
  JHEP {\bf 1208} (2012) 089
  [arXiv:1204.5977 [hep-ph]];
  %%CITATION = ARXIV:1204.5977;%% 
\bibitem{Badziak:2012rf}
  M.~Badziak, E.~Dudas, M.~Olechowski and S.~Pokorski,
  %``Inverted Sfermion Mass Hierarchy and the Higgs Boson Mass in the MSSM,''
  JHEP {\bf 1207} (2012) 155
  [arXiv:1205.1675 [hep-ph]].
  %%CITATION = ARXIV:1205.1675;%%
   
\bibitem{DLK} 
  M.~Dine, R.~G.~Leigh and A.~Kagan,
  %``Flavor symmetries and the problem of squark degeneracy,''
  Phys.\ Rev.\ D {\bf 48}, 4269 (1993)
  [hep-ph/9304299].
  %%CITATION = HEP-PH/9304299;%%
  %231 citations counted in INSPIRE as of 25 Jun 2013
  
\bibitem{PT} 
  A.~Pomarol and D.~Tommasini,
  %``Horizontal symmetries for the supersymmetric flavor problem,''
  Nucl.\ Phys.\ B {\bf 466}, 3 (1996)
  [hep-ph/9507462].
  %%CITATION = HEP-PH/9507462;%%
  %332 citations counted in INSPIRE as of 25 Jun 2013
  
 

\bibitem{fn}
  C.~D.~Froggatt and H.~B.~Nielsen,
  %``Hierarchy of Quark Masses, Cabibbo Angles and CP Violation,''
  Nucl.\ Phys.\ B {\bf 147} (1979) 277.
  %%CITATION = NUPHA,B147,277;%%

\bibitem{u1}
 M.~Leurer, Y.~Nir and N.~Seiberg,
  %``Mass matrix models,''
  Nucl.\ Phys.\ B {\bf 398} (1993) 319
  [hep-ph/9212278] and
  %%CITATION = HEP-PH/9212278;%%
 %``Mass matrix models: The Sequel,''
  Nucl.\ Phys.\ B {\bf 420} (1994) 468
  [hep-ph/9310320];
  %%CITATION = HEP-PH/9310320;%%
  
\bibitem{Dudas}
 E.~Dudas, S.~Pokorski and C.~A.~Savoy,
  %``Soft scalar masses in supergravity with horizontal U(1)-x gauge symmetry,''
  Phys.\ Lett.\ B {\bf 369} (1996) 255
  [hep-ph/9509410];
  %%CITATION = HEP-PH/9509410;%%
Y.~Kawamura and T.~Kobayashi,
  %``Soft scalar masses in string models with anomalous U(1) symmetry,''
  Phys.\ Lett.\ B {\bf 375} (1996) 141
   [Erratum-ibid.\ B {\bf 388} (1996) 867]
  [hep-ph/9601365];
  %%CITATION = HEP-PH/9601365;%%
 E.~Dudas, C.~Grojean, S.~Pokorski and C.~A.~Savoy,
  %``Abelian flavor symmetries in supersymmetric models,''
  Nucl.\ Phys.\ B {\bf 481} (1996) 85
  [hep-ph/9606383].
  %%CITATION = HEP-PH/9606383;%%  
  

\bibitem{Nelson}
A.~ E.~ Nelson, D.~ Wright,
Phys.Rev. D56 (1997) 1598-1604
 hep-ph/9702359. 

\bibitem{Chankowski:2005qp}
  P.~H.~Chankowski, K.~Kowalska, S.~Lavignac and S.~Pokorski,
  %``Update on fermion mass models with an anomalous horizontal U(1) symmetry,''
  Phys.\ Rev.\ D {\bf 71} (2005) 055004
  [hep-ph/0501071];
  %%CITATION = HEP-PH/0501071;%%
  %30 citations counted in INSPIRE as of 19 Jun 2013
  P.~H.~Chankowski, K.~Kowalska, S.~Lavignac and S.~Pokorski,	
 % Flavor changing neutral currents and inverted sfermion mass hierarchy
  [hep-ph/0507133 ].

\bibitem{Degenerate}
  Y.~Nir and N.~Seiberg,
  %``Should squarks be degenerate?,''
  Phys.\ Lett.\ B {\bf 309} (1993) 337
  [hep-ph/9304307].
  %%CITATION = HEP-PH/9304307;%%
  %428 citations counted in INSPIRE as of 03 Sep 2013
  
\bibitem{Perez2}
 O.~Gedalia, J.~F.~Kamenik, Z.~Ligeti and G.~Perez,
  %``On the Universality of CP Violation in Delta F = 1 Processes,''
  Phys.\ Lett.\ B {\bf 714} (2012) 55
  [arXiv:1202.5038 [hep-ph]].
  %%CITATION = ARXIV:1202.5038;%%
  %13 citations counted in INSPIRE as of 03 Sep 2013

\bibitem{BDH} 
  R.~Barbieri, G.~R.~Dvali and L.~J.~Hall,
  %``Predictions from a U(2) flavor symmetry in supersymmetric theories,''
  Phys.\ Lett.\ B {\bf 377}, 76 (1996)
  [hep-ph/9512388].
  %%CITATION = HEP-PH/9512388;%%
  %243 citations counted in INSPIRE as of 25 Jun 2013
  R.~Barbieri,  L.~J.~Hall and A.~Romanino,
  Phys.\ Lett.\ B {\bf 401}, 47 (1997)
  [hep-ph/9702315].
  
\bibitem{RRRS} 
  R.~G.~Roberts, A.~Romanino, G.~G.~Ross and L.~Velasco-Sevilla,
  %``Precision test of a fermion mass texture,''
  Nucl.\ Phys.\ B {\bf 615}, 358 (2001)
  [hep-ph/0104088].
  %%CITATION = HEP-PH/0104088;%%
  %105 citations counted in INSPIRE as of 25 Jun 2013

\bibitem{raby}
  R.~Dermisek and S.~Raby,
  %``Fermion masses and neutrino oscillations in SO(10) SUSY GUT with D(3) x U(1) family symmetry,''
  Phys.\ Rev.\ D {\bf 62} (2000) 015007
  [hep-ph/9911275].
  %%CITATION = HEP-PH/9911275;%%  
  
 \bibitem{BM} 
  K.~S.~Babu and R.~N.~Mohapatra,
  %``Supersymmetry, local horizontal unification, and a solution to the flavor puzzle,''
  Phys.\ Rev.\ Lett.\  {\bf 83}, 2522 (1999)
  [hep-ph/9906271].
  %%CITATION = HEP-PH/9906271;%%
  %31 citations counted in INSPIRE as of 25 Jun 2013 
  
\bibitem{heterotic}
 T.~Kobayashi, H.~P.~Nilles, F.~Ploger, S.~Raby and M.~Ratz,
  %``Stringy origin of non-Abelian discrete flavor symmetries,''
  Nucl.\ Phys.\ B {\bf 768} (2007) 135
  [hep-ph/0611020];
  %%CITATION = HEP-PH/0611020;%%
  H.~P.~Nilles, M.~Ratz and P.~K.~S.~Vaudrevange,
  %``Origin of Family Symmetries,''
  Fortsch.\ Phys.\  {\bf 61} (2013) 493
  [arXiv:1204.2206 [hep-ph]];
  %%CITATION = ARXIV:1204.2206;%%  
   M.~-C.~Chen, M.~Ratz and A.~Trautner,
  %``Non-Abelian discrete R symmetries,''
  arXiv:1306.5112 [hep-ph].
  
\bibitem{dbranes}
M.~Berasaluce-Gonzalez, L.~E.~Ibanez, P.~Soler and A.~M.~Uranga,
  %``Discrete gauge symmetries in D-brane models,''
  JHEP {\bf 1112} (2011) 113
  [arXiv:1106.4169 [hep-th]];
  %%CITATION = ARXIV:1106.4169;%%
 M.~Berasaluce-Gonzalez, P.~G.~Camara, F.~Marchesano, D.~Regalado and A.~M.~Uranga,
  %``Non-Abelian discrete gauge symmetries in 4d string models,''
  JHEP {\bf 1209} (2012) 059
  [arXiv:1206.2383 [hep-th]];
  %%CITATION = ARXIV:1206.2383;%%
  P.~Anastasopoulos, M.~Cvetic, R.~Richter and P.~K.~S.~Vaudrevange,
  %``String Constraints on Discrete Symmetries in MSSM Type II Quivers,''
  JHEP {\bf 1303} (2013) 011
  [arXiv:1211.1017 [hep-th]];
  %%CITATION = ARXIV:1211.1017;%%
   G.~Honecker and W.~Staessens,
  %``D6-Brane Model Building and Discrete Symmetries on T6/Z(2)xZ(6')xOR with Discrete Torsion,''
  arXiv:1303.6845 [hep-th];
F.~Marchesano, D.~Regalado and L.~Vazquez-Mercado,
  %``Discrete flavor symmetries in D-brane models,''
  arXiv:1306.1284 [hep-th].
  %%CITATION = ARXIV:1306.1284;%%  

\bibitem{ibanezross}
 L.~E.~Ibanez and G.~G.~Ross,
  %``Fermion masses and mixing angles from gauge symmetries,''
  Phys.\ Lett.\  B {\bf 332} (1994) 100
  [arXiv:hep-ph/9403338];
  %%CITATION = PHLTA,B332,100;%%
P.~Binetruy and P.~Ramond,
  %``Yukawa textures and anomalies,''
  Phys.\ Lett.\  B {\bf 350} (1995) 49
  [arXiv:hep-ph/9412385];
  %%CITATION = PHLTA,B350,49;%%
  E.~Dudas, S.~Pokorski and C.~A.~Savoy,
  %``Yukawa matrices from a spontaneously broken Abelian symmetry,''
  Phys.\ Lett.\  B {\bf 356} (1995) 45
  [arXiv:hep-ph/9504292]; Y.~Nir,
  %``Gauge unification, Yukawa hierarchy and the mu problem,''
  Phys.\ Lett.\  B {\bf 354} (1995) 107
  [arXiv:hep-ph/9504312];
  %%CITATION = PHLTA,B354,107;%%
 P.~Binetruy, S.~Lavignac and P.~Ramond,
  %``Yukawa textures with an anomalous horizontal Abelian symmetry,''
  Nucl.\ Phys.\ B {\bf 477} (1996) 353
  [hep-ph/9601243];
  %%CITATION = HEP-PH/9601243;%%  
 N.~Irges, S.~Lavignac and P.~Ramond,
  %``Predictions from an anomalous U(1) model of Yukawa hierarchies,''
  Phys.\ Rev.\ D {\bf 58} (1998) 035003
  [hep-ph/9802334];
  %%CITATION = HEP-PH/9802334;%%  
 H.~K.~Dreiner, H.~Murayama and M.~Thormeier,
  %``Anomalous flavor U(1)(X) for everything,''
  Nucl.\ Phys.\ B {\bf 729} (2005) 278
  [hep-ph/0312012].
  %%CITATION = HEP-PH/0312012;%%  

\bibitem{ibanez}
L.~E.~Ibanez,
  %``Computing the weak mixing angle from anomaly cancellation,''
  Phys.\ Lett.\ B {\bf 303} (1993) 55
  [hep-ph/9205234].
  %%CITATION = HEP-PH/9205234;%%

\bibitem{GS}
 M.~B.~Green and J.~H.~Schwarz,
  %``Anomaly Cancellation in Supersymmetric D=10 Gauge Theory and Superstring Theory,''
  Phys.\ Lett.\ B {\bf 149} (1984) 117.
  %%CITATION = PHLTA,B149,117;%%
  %2187 citations counted in INSPIRE as of 07 Dec 2013
  
\bibitem{Olechowski:1990bh}
  M.~Olechowski and S.~Pokorski,
  %``Heavy top quark and scale dependence of quark mixing,''
  Phys.\ Lett.\ B {\bf 257} (1991) 388.
  %%CITATION = PHLTA,B257,388;%%
  %76 citations counted in INSPIRE as of 24 Jun 2013
  
\bibitem{Beringer:1900zz}
  J.~Beringer {\it et al.}  [Particle Data Group Collaboration],
  %``Review of Particle Physics (RPP),''
  Phys.\ Rev.\ D {\bf 86} (2012) 010001.
  %%CITATION = PHRVA,D86,010001;%%
  %1651 citations counted in INSPIRE as of 24 Jun 2013

\bibitem{discrete}
  A.~Blum, C.~Hagedorn and M.~Lindner,
  %``Fermion Masses and Mixings from Dihedral Flavor Symmetries with Preserved Subgroups,''
  Phys.\ Rev.\ D {\bf 77} (2008) 076004
  [arXiv:0709.3450 [hep-ph]].
  %%CITATION = ARXIV:0709.3450;%%

\bibitem{Mescia}
  F.~Mescia and J.~Virto,
  %``Natural SUSY and Kaon Mixing in view of recent results from Lattice QCD,''
  Phys.\ Rev.\ D {\bf 86} (2012) 095004
  [arXiv:1208.0534 [hep-ph]].
  %%CITATION = ARXIV:1208.0534;%%
  %8 citations counted in INSPIRE as of 02 Sep 2013
  
\bibitem{UTfit} 
  V.~Bertone {\it et al.}  [ ETM Collaboration],
  %``Kaon Mixing Beyond the SM from Nf=2 tmQCD and model independent constraints from the UTA,''
  JHEP {\bf 1303} (2013) 089
  [arXiv:1207.1287 [hep-lat]].
  %%CITATION = ARXIV:1207.1287;%%
  %17 citations counted in INSPIRE as of 17 Jul 2013
  
 \bibitem{hagelin}
  J.~S.~Hagelin, S.~Kelley and T.~Tanaka,
  %``Supersymmetric flavor changing neutral currents: Exact amplitudes and phenomenological analysis,''
  Nucl.\ Phys.\ B {\bf 415} (1994) 293;
  %%CITATION = NUPHA,B415,293;%%
 F.~Gabbiani, E.~Gabrielli, A.~Masiero and L.~Silvestrini,
  %``A Complete analysis of FCNC and CP constraints in general SUSY extensions of the standard model,''
  Nucl.\ Phys.\ B {\bf 477} (1996) 321
  [hep-ph/9604387].
  %%CITATION = HEP-PH/9604387;%%

\bibitem{Lalak}
 Z.~ Lalak, S.~ Pokorski and  G.~ G. ~Ross,
JHEP 1008 (2010) 129
arXiv:1006.2375 [hep-ph].

\bibitem{GNR} 
  G.~F.~Giudice, M.~Nardecchia and A.~Romanino,
  %``Hierarchical Soft Terms and Flavor Physics,''
  Nucl.\ Phys.\ B {\bf 813}, 156 (2009)
  [arXiv:0812.3610 [hep-ph]].
  %%CITATION = ARXIV:0812.3610;%%



\bibitem{Hisano}
  J.~Hisano, T.~Moroi, K.~Tobe and M.~Yamaguchi,
  %``Lepton flavor violation via right-handed neutrino Yukawa couplings in supersymmetric standard model,''
  Phys.\ Rev.\ D {\bf 53} (1996) 2442
  [hep-ph/9510309].
  %%CITATION = HEP-PH/9510309;%%
  %552 citations counted in INSPIRE as of 02 Aug 2013
    
\bibitem{MEG}
 J.~Adam {\it et al.}  [MEG Collaboration],
  %``New constraint on the existence of the mu+-> e+ gamma decay,''
  arXiv:1303.0754 [hep-ex].
  %%CITATION = ARXIV:1303.0754;%%
  %39 citations counted in INSPIRE as of 02 Sep 2013

\bibitem{Allanach:2001kg}
  B.~C.~Allanach,
  %``SOFTSUSY: a program for calculating supersymmetric spectra,''
  Comput.\ Phys.\ Commun.\  {\bf 143} (2002) 305
  [hep-ph/0104145].

\bibitem{Rosiek}
 J.~Rosiek,
Phys.Rev. D41 (1990) 3464
Erratum: hep-ph/9511250.
  %%CITATION = HEP-PH/9511250;%%
  %138 citations counted in INSPIRE as of 15 May 2013
  
 
  


%  M.~Bona {\it et al.}  [UTfit Collaboration],
  %``Model-independent constraints on $\Delta$ F=2 operators and the scale of new physics,''
%  JHEP {\bf 0803}, 049 (2008)
%  [arXiv:0707.0636 [hep-ph]].
  %%CITATION = ARXIV:0707.0636;%%
  %263 citations counted in INSPIRE as of 31 May 2013
%\bibitem{Bertone:2012cu}


\end{thebibliography}
\end{document}